\UseRawInputEncoding
\documentclass[%
 reprint,
amsmath,amssymb,
aps,
prx,
longbibliography,
]{revtex4-2}

\usepackage{graphicx}% Include figure files
\usepackage{dcolumn}% Align table columns on decimal point
\usepackage{bm}% bold math
\usepackage{hyperref}
\usepackage{xcolor}
\begin{document}

% \preprint{APS/123-QED}

\title{King-plot analysis of isotope shifts in simple diatomic molecules}

\author{Michail Athanasakis-Kaklamanakis}
%  \altaffiliation[Also at ]{Physics Department, XYZ University.}%Lines break automatically or can be forced with \\
 \email{m.athkak@cern.ch}
\affiliation{%
 Experimental Physics Department, CERN, CH-1211 Geneva 23, Switzerland\\
 KU Leuven, Instituut voor Kern- en Stralingsfysica, B-3001 Leuven, Belgium
}%
% \collaboration{CRIS Collaboration at CERN-ISOLDE}%\noaffiliation

\author{Shane G. Wilkins}
\affiliation{%
 Department of Physics, Massachusetts Institute of Technology, Cambridge, MA 02139, USA \\
 Laboratory for Nuclear Science, Massachusetts Institute of Technology, Cambridge, MA 02139, USA
}%

\author{Alexander A. Breier}
\affiliation{Laboratory for Astrophysics, Institute of Physics, University of Kassel, 34132 Kassel, Germany}

\author{Gerda Neyens}
% \email{gerda.neyens@kuleuven.be}
\affiliation{
  KU Leuven, Instituut voor Kern- en Stralingsfysica, B-3001 Leuven, Belgium
}%

\date{\today}% It is always \today, today,

\begin{abstract}
We demonstrate that the isotope shift in isotopomers of diatomic molecules, where the nucleus of one of its constituent atoms is replaced by another isotope, can be expressed as the sum of a field shift and a mass shift, similar to the atomic case. We show that a linear relation holds between atomic and molecular isotopes shifts, thus extending the King-plot analysis to molecular isotope shifts. Optical isotope shifts in YbF and ZrO and infrared isotope shifts in SnH are analyzed with a molecular King-plot approach, utilizing Yb$^{+}$ and Zr$^{+}$ ionic isotope shifts and charge radii of Sn obtained with non-optical methods. The changes in the mean-squared nuclear charge radii~$\delta \langle r^2 \rangle$ of $^{170-174,176}$Yb and $^{90-92,94,96}$Zr extracted from the molecular transitions are found to be in excellent agreement with the values from the spectroscopy of Yb$^{+}$ and Zr$^{+}$, respectively. On the contrary, in the case of the vibrational-rotational transition in SnH, no sensitivity to the nuclear volume could be deduced within the experimental resolution, which makes it unsuitable for the extraction of nuclear charge radii but provides insights into the molecular electronic wavefunction not accessible via other methods. The new opportunities offered by the molecular King-plot analysis for research in nuclear structure and molecular physics is discussed.
\end{abstract}

%Use showkeys class option if keyword
\maketitle

%\tableofcontents

\section{\label{sec:intro}Introduction}
% \subsection{Laser spectroscopy for nuclear structure studies}
% \subsection{Isotope-shift spectroscopy in atoms}
When the same atomic transition is measured with high-enough precision in atoms (or ions) containing two different isotopes of the same element, a small change in the transition frequency is observed, called the \textit{isotope shift}, being the sum of the \textit{field shift} and the \textit{mass shift}~\cite{King2013}:
\begin{equation}
    \delta \nu_{\mathrm{IS}}^{A,A'} = \delta \nu_{\mathrm{FS}}^{A,A'} + \delta \nu_{\mathrm{MS}}^{A,A'}
\end{equation}
% The term $\delta \nu_{IS}^{A,A'}$ is the isotope shift in the transition frequency between an isotope $A$ and the reference isotope $A'$ and arises from the difference in the nuclear volume and mass between the two isotopes. For atoms, it is defined as:

The field shift (FS)~$\delta \nu_{\mathrm{FS}}^{A,A'}$ originates from the change in the electromagnetic field experienced by core-penetrating electrons and is thus linked to the change in the mean-squared nuclear charge radius~$\delta \langle r^2 \rangle^{A,A'}$, which gives information about the proton distribution in the nucleus. For this reason, isotope shifts have been measured for a large number of elements and isotopes, both stable and radioactive, as they provide key information on the evolution of nuclear charge radii across the nuclear chart~\cite{Campbell2016}. 
% As a nuclear-structure observable of rising interest, the 
The nuclear charge radius is being actively studied as a quantity through which a number of unique nuclear phenomena can manifest, and whose systematic study can provide direct insights into the strong nuclear interaction and shell structure~\cite{Reinhard2020,Reinhard2021,Koszorus2021}.
% The mass shift $\delta \nu_{\mathrm{MS}}^{A,A'}$ arises from the change in the center of atomic mass (normal mass shift, NMS) and the electron correlation energy (specific mass shift, SMS). 

The mass shift $\delta \nu_{\mathrm{MS}}^{A,A'}$, on the other hand, arises from the difference in mass between the two isotopes and has two contributions: a shift due to the slight change to the center of nuclear mass with respect to the electrons (normal mass shift, NMS) and a shift due to the change in the correlation energy between electrons (specific mass shift, SMS).
Overall, the isotope shift can be expressed in terms of the difference in the mean-squared nuclear charge radii of the two isotopes and the relative difference of their atomic masses:
\begin{eqnarray}
    % \nonumber
    % \delta \nu_{IS}^{A,A'} = \delta \nu_{FS}^{A,A'} + \delta \nu_{MS}^{A,A'}
    % \\
    \nonumber
    \delta \nu_{\mathrm{FS}}^{A,A'} = F \delta \langle r^2 \rangle ^{A,A'}
    \\\nonumber
    \delta \nu_{\mathrm{MS}}^{A,A'} = (K_{NMS}+K_{SMS})\frac{M_{A'}-M_A}{M_{A'}M_A}
\end{eqnarray}
\vspace{-13pt}
\begin{equation} \label{eqn:atomic_is}
    \therefore \hspace{3.5pt} \delta \nu_{\mathrm{IS}}^{A,A'} = F \delta \langle r^2 \rangle ^{A,A'} + K \frac{M_{A'}-M_A}{M_{A'}M_A}
\end{equation}
where $F$ and $K$ are called the \textit{field-shift} and \textit{mass-shift} factors, respectively and are specific to the electronic transition that is studied. While the NMS factor $K_{NMS}$ can be calculated exactly due to its classical form, the field-shift factor \textit{F} and the SMS factor $K_{SMS}$ require atomic-structure calculations that include two-body interactions between all electrons~\cite{King2013}. Such calculations can be challenging or impossible with state-of-the-art methods, depending on the atomic structure.

%
% By measuring the hyperfine splitting and isotope shift in the transitions of atomic electrons, laser spectroscopy provides access to the electromagnetic moments and mean-squared charge radii of nuclei across long isotopic chains, and has thus become a well-established technique at RIB facilities to study the structure of radioactive nuclei with half-lives down to a few milliseconds.

% In the last decade, a new area of experimental research has also emerged, using high-precision laser spectroscopy across isotopic chains to study the potential signature of physics beyond the Standard Model~\cite{Counts2020,Solaro2020,Miyake2019,Knollmann2019}. Such studies utilize the linearity of Eqn.~\ref{eqn:atomic_is} in the form of a King-plot analysis.

To overcome this problem, the atomic factors for a transition $i$ that cannot be accurately calculated are often derived from the calculated ones in a transition $j$ using a King-plot analysis~\cite{King2013}.  It makes use of the linear relationship between the isotope shifts in two transitions in the same element as:
\begin{eqnarray}
    \nonumber
    \tilde{M}^{A,A'} \delta \nu_{i}^{A,A'} = \tilde{M}^{A,A'} F_i \delta \langle r^2 \rangle ^{A,A'} + K_i
    \\ \nonumber
    \tilde{M}^{A,A'} \delta \nu_{j}^{A,A'} = \tilde{M}^{A,A'} F_j \delta \langle r^2 \rangle ^{A,A'} + K_j
\end{eqnarray}
\vspace{-19pt}
\begin{equation} \label{eqn:kingplot_atom}
    \therefore \hspace{3.5pt} \tilde{M}^{A,A'} \delta \nu_{i}^{A,A'} = \frac{F_i}{F_j} \tilde{M}^{A,A'} \delta \nu_{j}^{A,A'} + K_i - \frac{F_i}{F_j} K_j
\end{equation}
where $\tilde{M}^{A,A'}= \frac{M_{A'}M_A}{M_{A'}-M_A}$. The relationship between $\tilde{M}^{A,A'} \delta \nu_{i}^{A,A'}$ and $ \tilde{M}^{A,A'} \delta \nu_{j}^{A,A'}$ in Eqn.~\ref{eqn:kingplot_atom} is linear and therefore, in a plot of $\tilde{M}^{A,A'} \delta \nu_{i}^{A,A'}$ against $ \tilde{M}^{A,A'} \delta \nu_{j}^{A,A'}$, the slope and y-intercept of the linear fit will relate the isotope-shift factors of the two transitions~\cite{King2013}. Importantly, the King-plot analysis can also be applied to transitions in different charge states of the same atom~(for example, see Ref.~\cite{Vormawah2018}), allowing further optimization of the experimental conditions, such as favorable chemistry.

Due to the relation between the isotope shift and the difference in mean-squared charge radii of the two isotopes, isotope-shift measurements with high-precision atomic spectroscopy have been a cornerstone of nuclear structure studies with laser spectroscopy at radioactive ion beam (RIB) facilities~\cite{Blaum2013,Campbell2016}.

By tracing the evolution of the isotope shift across an isotopic range, the changes in mean-squared nuclear charge radii can be studied with high accuracy. The nuclear charge radius is a nuclear-structure observable of importance, as precision measurements across the nuclear chart continue to challenge the predictive ability of \textit{ab initio} nuclear theory~\cite{GarciaRuiz2016,deGroote2020b,Koszorus2021,Kortelainen2021}, and by extension our understanding of the observable. Such precision studies in nuclei with extreme proton-to-neutron ratios are a key pathway to obtaining new insights into the strong nuclear interaction.

While in previous decades, models of the nucleon-nucleon interaction used to be derived empirically, progress in nuclear theory has nowadays led to \textit{ab initio} derivations from quantum chromodynamics using chiral effective field theory~\cite{Hammer2020}. In combination with modern methods to solve the many-body Schr\"{o}dinger equation and growing computational power~\cite{Barrett2013,Hagen2014,Navratil2016,Hergert2016,Soma2021}, \textit{ab initio} calculations of nuclear observables are now available not only for light nuclei, where rare nuclear phenomena such as nucleon superfluidity~\cite{Miller2019}, nucleon clustering and bubble structures~\cite{Chernykh2007,KanadaEnyo2015,Duguet2017}, many-body currents and continuum effects~\cite{Hagen2012,Carlson2015}, and many-body nuclear forces~\cite{Hagen2012,Ekstroem2015,Novario2020} have been investigated, but also for nuclei as heavy as $^{208}$Pb~\cite{Hu2022}.

However, extending the measurements of nuclear charge radii to more weakly bound isotopes or to new isotopic chains is often challenging due to the atomic structure of the species of interest~\cite{Campbell2016}. For example, the high reactivity and the predominance of extreme-ultraviolet electronic transitions from the ground state have hindered the study of most atomic chains in the oxygen region, while many heavier nuclei form refractory atoms, preventing their production and study at traditional RIB facilities~\cite{Campbell2016}.

In several cases, studying the nucleus as part of a diatomic molecule could provide a solution to this problem. In the past, radioactive molecular beams containing a reactive or refractory element have been observed at ISOL facilities, including as a method to extract a particular element more easily from the target and/or to separate it from its isobars more effectively~\cite{Koester2008,Ballof2021Thesis}. Recently, laser-spectroscopic studies of short-lived radioactive molecules were performed for the first time~\cite{GarciaRuiz2020,Udrescu2021} and the potential benefit of studying the spectra of radioactive molecules for a number of scientific areas has also been recognized~\cite{Athanasakis2021}, such as astrophysics, medical-isotope production, fundamental physics, and nuclear structure. To be beneficial for nuclear structure studies, molecular laser spectroscopy should be capable of providing access to the nuclear moments and changes in mean-squared charge radii with a comparable or higher precision compared to atomic spectroscopy. While the study of the molecular hyperfine structure with laser spectroscopy is well-established for molecules containing stable isotopes~\cite{Demtroder1982,Truppe2019,Pilgram2021}, the influence of nuclear size effects on the molecular isotope shifts has received less attention. Early works on molecular isotope shifts focused on the presence of the mass shift~\cite{Ross1974, Watson1980}, with later studies also establishing the presence of a field shift in the isotope shifts in PbS and PbO~\cite{Knoeckel1982,Knoeckel1984, Knoeckel1985}. Recently, the first optical isotope shifts in short-lived isotopomers of a molecule were reported for RaF~\cite{Udrescu2021}. However, in all molecular studies so far, measurements of $\delta \langle r^2 \rangle^{A,A'}$ from atomic laser spectroscopy have been used as input in the analysis, to compare the extracted electronic density around the nuclear volume with computational predictions. To the knowledge of the authors, no study so far has focused on $\delta \langle r^2 \rangle^{A,A'}$ as an observable to be extracted from the molecular isotope shifts for cases where a value is not already available from other methods. One of the reasons is that the accuracy of mass-shift calculations in molecules has not been sufficiently benchmarked. Therefore, any experimental progress in the study of nuclear size effects in molecules would require extensive theoretical work.

In this work, a different approach is proposed: it is demonstrated that the King-plot analysis framework for atomic spectroscopy can be extended to diatomic molecules. This is demonstrated here for simple diatomic molecules having a $\Sigma$ electronic ground state with few valence electrons. However, the presented formalism to express the molecular isotopomer shift as the sum of a field and a mass shift where only one of the two atoms is modified is generally applicable. By then relating the well-studied atomic field- and mass-shift factors with the molecular counterparts, rapid experimental progress in studying size effects in unstable nuclei with molecular laser spectroscopy could be enabled. The validity of the method is tested, as first examples, with literature optical isotope-shift measurements in YbF and ZrO, for Yb and Zr containing multiple stable isotopes. It is shown that the values of $\delta \langle r^2 \rangle^{A,A'}$ extracted from the molecular isotope shifts following a King-plot analysis with atomic (ionic) data are consistent with the charge radii extracted from atomic (ionic) spectroscopy. The molecular King-plot approach is also explored for vibrational-rotational transitions in SnH. From the analysis of YbF, ZrO, and SnH, it is observed that the molecular King-plot analysis can also be a sensitive tool to study the electronic wavefunction in diatomic molecules, providing a versatile tool to obtain molecular-structure insights.
% Sections~\ref{sec:derivation} and \ref{sec:analysis} are discussed in the contexts of nuclear and molecular physics research in Section~\ref{sec:discussion_outlook}, also highlighting new opportunities.
%
%
%
%
\section{\label{sec:derivation}The molecular isotope shift}
\begin{figure}[h]
    \centering
    \includegraphics[width=8.6cm]{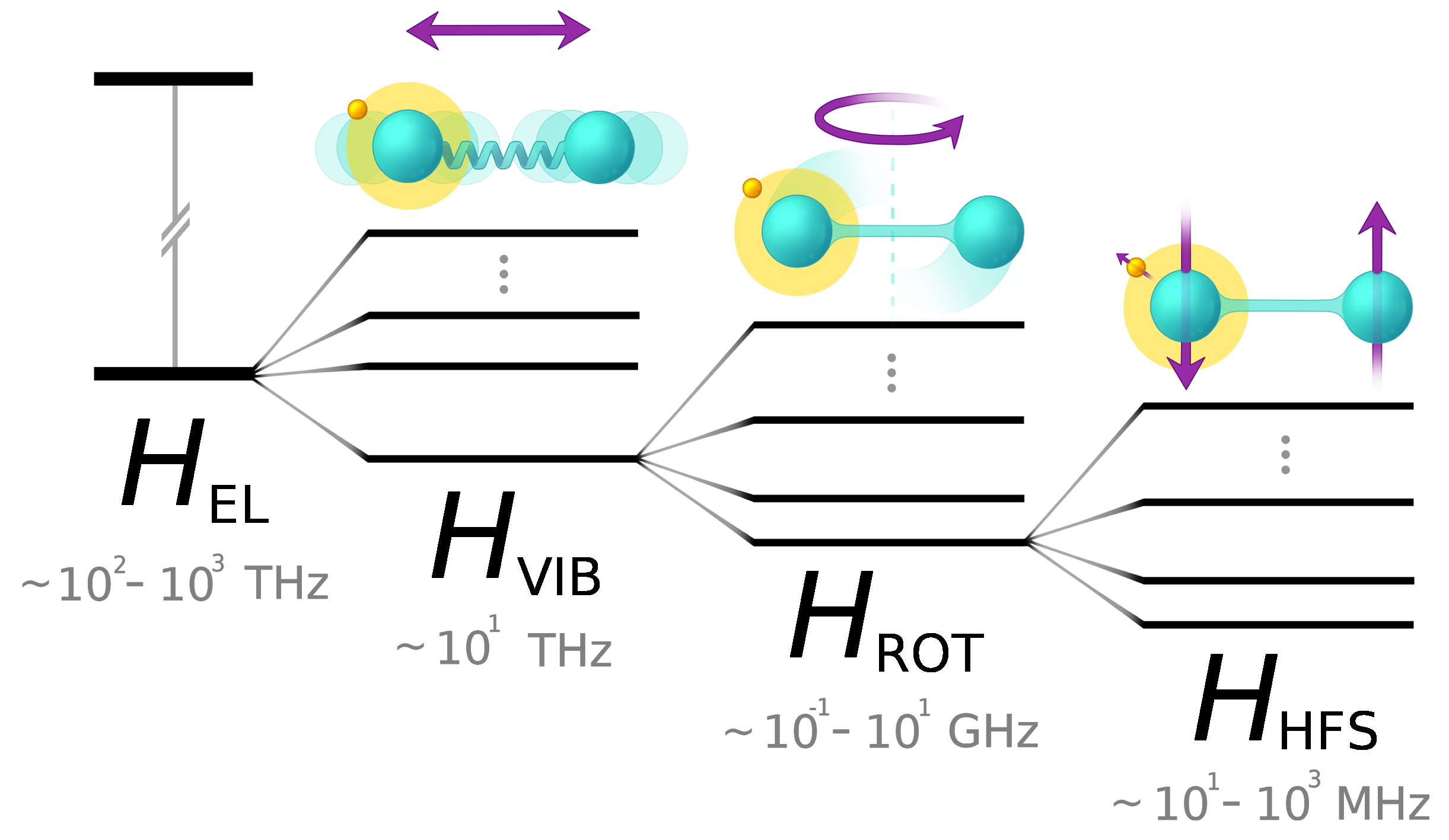}
    \caption{\small Vibrational, rotational, and hyperfine structures of an electronic state in a simple diatomic molecule. The substructure splittings are schematic and are not representative of physical scales, but typical frequency scales are given under each term.}
    \label{fig:molecular_levels}
\end{figure}

% The linear form for the isotope shift in a rovibronic transition in a diatomic molecule can be derived, including all higher-order corrections in the Dunham expansion.
% \\\\
Molecular electronic states possess vibrational and rotational substructure, typically leading to very rich spectra. In simplistic terms, for each electronic state in a diatomic molecule, an associated series of energy levels due to vibrations of the atomic bond is added on top of the electronic energy, with further, smaller energy contributions to each vibrational substate stemming from the rotational energy of the molecule. Further splittings due to the fine and the hyperfine structure are also present. The simplified description of the rovibronic (rotational, vibrational, and electronic) and hyperfine structure is shown pictorially in Fig.~\ref{fig:molecular_levels}.

The energy of a diatomic molecular level can be parametrized according to the Dunham expansion \cite{Dunham1932}:
\begin{eqnarray}
\label{eqn:dunham} E_{\nu, J}^{\Lambda} = \sum_{(k,l) \geq (0,0)} Y_{kl}^\Lambda  \left ( \nu + \frac{1}{2}\right )^k \left [J(J+1)\right ]^l
\end{eqnarray}
where the $\hat{N}^2$ notation is used in the present work, thus $\Lambda$ denotes some electronic state and $\nu$ and $J$ are the vibrational and rotational quantum numbers, respectively. $Y_{kl}$ are the Dunham parameters~\footnote{Not to be confused with the spherical tensors, that are typically represented by the same symbol.} and they are related to the usual molecular parameters, such that $Y_{00}$ corresponds to the electronic term energy $T_e$, $Y_{10}$ corresponds to the harmonic frequency $\omega_e$, $Y_{01}$ to the rotational constant $B_e$, etc. The powers $k$ and $l$ quantify the order of vibrational and rotational correction to the energy, respectively.

Almost a century ago, it was noticed that the ratio of the values of a Dunham parameter in two isotopomers~$X^{\prime}$ and $X^{\prime\prime}$ of the same molecule is \emph{almost} equal to the ratio of their reduced masses, raised to the power of $l+k/2$. Due to the presence of higher-order corrections, however, the equivalence is not exact~\cite{Dunham1932,Herzberg1945}:
\begin{equation*}
    \frac{Y^{\Lambda,X^{\prime}}_{kl}}{Y^{\Lambda,X^{\prime\prime}}_{kl}} \approx \bigg( \frac{\mu_{X^{\prime\prime}}}{\mu_{X^{\prime}}} \bigg) ^{l+k/2}
\end{equation*}
where $\mu = \frac{M_A M_B}{M_A + M_B}$ is the reduced mass of the diatomic molecule with $A$ and $B$ being its constituent atoms.

The ratio can be made exact with the introduction of a parameter~$U_{kl}^\Lambda$ that groups the higher-order corrections in the Dunham parameters~\cite{Knecht2012,Almoukhalalati2016} as:
\begin{eqnarray}
Y_{kl}^{\Lambda} = \mu ^{-(l+k/2)} U_{kl}^\Lambda
\end{eqnarray}
and so equation \ref{eqn:dunham} can be written as:
\begin{eqnarray}
\label{eqn:dunhammass} E_{\nu, J}^{\Lambda} = \sum_{(k,l) \geq (0,0)} \mu ^{-(l+k/2)} U_{kl}^\Lambda  \left ( \nu + \frac{1}{2}\right )^k \left [J(J+1)\right ]^l
\end{eqnarray}

Additionally, $U_{kl}^\Lambda$ can be further decomposed into an isotope-independent parameter, $\tilde{U}_{kl}^\Lambda$, and terms that depend on the masses and the extended nuclear volumes of the atomic constituents~\cite{Schlembach1982}:
% (see Ref.~\cite{Udrescu2021} and references within):
% \begin{eqnarray}
% \begin{split}
% U_{kl}^\Lambda = \tilde{U}_{kl}^\Lambda \left [1 + m_e \left( \frac{\Delta_{kl}^{\Lambda, A}}{M_A} + \frac{\Delta_{kl}^{\Lambda, B}}{M_B} \right) \right. \\
% + \left. V_{kl}^{\Lambda,A} \left \langle r^2 \right \rangle _A + V_{kl}^{\Lambda,B} \left \langle r^2 \right \rangle _B  \right]
% \end{split}
% \end{eqnarray}
% 
\begin{multline}
U_{kl}^\Lambda = \tilde{U}_{kl}^\Lambda \left [1 + m_e \left( \frac{\Delta_{kl}^{\Lambda, A}}{M_A} + \frac{\Delta_{kl}^{\Lambda, B}}{M_B} \right) \right. \\
+ \left. V_{kl}^{\Lambda,A} \left \langle r^2 \right \rangle _A + V_{kl}^{\Lambda,B} \left \langle r^2 \right \rangle _B  \right]
\end{multline}
where $m_e$ is the electron mass, $M_A$ and $M_B$ are the masses of atoms $A$ and $B$ respectively, and $\langle r^2 \rangle$ is the mean-squared nuclear charge radius.
% The letters \textit{A} and \textit{B} as superscripts indicate a dependence on the atom~\textit{A} and atom~\textit{B} of the diatomic molecule, while as subscripts they denote a property of atom~\textit{A} and atom~\textit{B}, respectively.

The parameters $\Delta_{kl}^{\Lambda, A}$ and $\Delta_{kl}^{\Lambda, B}$ are specific to each atom~$A$ and $B$ in the diatomic molecule but invariant under isotopic substitution, and they quantify the sensitivity of the electronic, vibrational, and rotational (hence rovibronic) state energy to the corresponding atomic mass, in the form of corrections to the Born-Oppenheimer approximation. Similarly, the isotope-independent parameters $V_{kl}^{\Lambda,A}$ and $V_{kl}^{\Lambda,B}$ quantify the sensitivity of the state energy to the extended volume of the corresponding atomic nucleus, and they are independent of the nuclear mass.

Therefore, overall, the energy of a rovibronic state~$\Lambda$ in a diatomic molecule can be expressed with explicit dependence on the atomic masses and the extended nuclear volumes of the constituent atoms as:
% \begin{eqnarray}
% \begin{split}
% \label{eqn:dunhamfullold}
% E_{\nu, J}^{\Lambda} = \sum_{k,l \geq 0} \mu ^{-(l+k/2)} \left ( \nu + \frac{1}{2}\right )^k \left [J(J+1)\right ]^l \\
% U_{kl}^\Lambda \Bigg[ 1 + m_e \Bigg( \frac{\Delta_{kl}^{\Lambda, A}}{M_A} + \frac{\Delta_{kl}^{\Lambda, B}}{M_B} \Bigg)  \\
% + V_{kl}^{\Lambda,A} \left \langle r^2 \right \rangle _A + V_{kl}^{\Lambda,B} \left \langle r^2 \right \rangle _B  \Bigg]
% \end{split}
% \end{eqnarray}
\begin{multline}
\label{eqn:dunhamfullold}
E_{\nu, J}^{\Lambda} = \sum_{(k,l) \geq (0,0)} \mu ^{-(l+k/2)} \left ( \nu + \frac{1}{2}\right )^k \left [J(J+1)\right ]^l \tilde{U}_{kl}^\Lambda 
\displaybreak[0]\\
\times \Bigg[ 1 + m_e \Bigg( \frac{\Delta_{kl}^{\Lambda, A}}{M_A} + \frac{\Delta_{kl}^{\Lambda, B}}{M_B} \Bigg)  
+ V_{kl}^{\Lambda,A} \left \langle r^2 \right \rangle _A + V_{kl}^{\Lambda,B} \left \langle r^2 \right \rangle _B  \Bigg]
\end{multline}

A detailed discussion and definition for the field parameters $V^{\Lambda, A/B}_{kl}$  is given in Ref.~\cite{Almoukhalalati2016} (in this text, $A/B$ denotes \textit{"A or B"} and not a numerical ratio). The field parameters are dependent on the nuclear charge~$Z^{\alpha/\beta}$ of atom~$A/B$ containing isotope~$\alpha/\beta$, the fundamental charge~$e$, the vacuum permittivity~$\epsilon_0$, the molecular force constant~$k_e$, the electron density~$E_{el}^{\Lambda,A/B}$ at the nucleus of $A/B$ for state $\Lambda$, the equilibrium internuclear distance~$R_e$, and the second expansion coefficient of the Dunham potential~$\alpha_1$~\cite{Almoukhalalati2016}. Almoukhalalati \textit{et al.}~\cite{Almoukhalalati2016} demonstrated that, while further improvements are required to accurately match the values fitted from experiment, calculations of the field parameters are tractable with \textit{ab initio} theory and the parameter definitions derived in their work are generally accurate.

On the other hand, the mass-related parameters~$\Delta ^{\Lambda, A/B}_{kl}$ are defined by Eqn.~\ref{eqn:dunhamfullold}~\cite{Ross1974} and, although Watson provided theoretical justification for the addition of these parameters~\cite{Watson1980}, several problems with the formulation of Eqn.~\ref{eqn:dunhamfullold} were later identified by Le Roy~\cite{LeRoy1999}. Instead, it was proposed~\cite{LeRoy1999} that a multi-isotopomer analysis in molecules is used instead, which incorporates isotope referencing in the Dunham expansion as:
\begin{multline}
\label{eqn:dunhamfull}
    E^{\Lambda, X^{\prime}}_{\nu, J} = 
    \displaybreak[0]\\
    \sum_{(k,l) \neq (0,0)}  \Big(  \frac{\mu_{X^{\prime\prime}}}{\mu_{X^\prime}} \Big)^{l+k/2} \Big(\nu + \frac{1}{2} \Big)^{k} \big[ J(J+1)\big]^{l} Y^{\Lambda,X^{\prime\prime}}_{kl}
    \displaybreak[0]\\
    + \sum_{(k,l) \geq (0,0)}  \Big(  \frac{\mu_{X^{\prime\prime}}}{\mu_{X^\prime}} \Big)^{l+k/2} \Big(\nu + \frac{1}{2} \Big)^{k} \big[ J(J+1)\big]^{l} 
    \displaybreak[0]\\
    \times \Bigg( \frac{\Delta M^{X^\prime}_{A}}{M^{X^\prime}_{A}}\delta^{\Lambda,A}_{kl} + \frac{\Delta M^{X^\prime}_{B}}{M^{X^\prime}_{B}}\delta^{\Lambda,B}_{kl} 
    \displaybreak[0]\\
    + \delta \langle r^2 \rangle ^{X^\prime}_{A} f^{\Lambda, A}_{kl} + \delta \langle r^2 \rangle ^{X^\prime}_{B} f^{\Lambda, B}_{kl} \Bigg)
\end{multline}
where $X$ is a diatomic molecule containing atoms $A$ and $B$, $X^{\prime}$ is some isotopomer with isotopes $\alpha^{\prime}$ and $\beta^{\prime}$ for the two atoms respectively, $X^{\prime\prime}$ is the reference isotopomer with isotopes $\alpha^{\prime\prime}$ and $\beta^{\prime\prime}$, $\Delta M^{X^\prime}_{A}=M^{\alpha^{\prime\prime}}_{A} - M^{\alpha^{\prime}}_{A}$ is the difference in atomic mass between the reference isotope and the isotope in question for atom $A$, and $\delta \langle r^2 \rangle ^{X^{\prime}}_{A} = \langle r^2 \rangle_{\alpha^{\prime}} - \langle r^2 \rangle_{\alpha^{\prime\prime}}$ is the change in the mean-squared charge radius of the nucleus in atom~$A$. Identical expressions for $\Delta M^{X^\prime}_{A}$ and $\delta \langle r^2 \rangle ^{X^{\prime}}_{A}$ also hold for atom~$B$. As seen in Eqn.~\ref{eqn:dunhamfull}, the energy of the rovibronic state~$\Lambda$ in isotopomer $X^{\prime}$ is expressed as a series of isotopomer-specific corrections to the Dunham parameters of the reference isotopomer~$X^{\prime\prime}$.

The modified mass parameters $\delta^{\Lambda,A/B}_{kl}$ introduced by Le Roy are related to the mass parameters~$\Delta ^{\Lambda, A/B}_{kl}$ defined by Ross et al.~\cite{Ross1974} (the relation is described in Ref.~\cite{LeRoy1999}) while preventing the complications created by the latter. The modified field parameters $f^{\Lambda, A/B}_{kl}$ are also related to the volume parameters analyzed by Almoukhalalati et al.~\cite{Almoukhalalati2016} in a similar fashion. As the Le~Roy formalism is fully equivalent with that of Eqn.~\ref{eqn:dunhamfullold}~\cite{LeRoy1999}, the work of Almoukhalalati et al.~\cite{Almoukhalalati2016}, which is based on Eqn.~\ref{eqn:dunhamfullold}, is also fully consistent with Eqn.~\ref{eqn:dunhamfull} and their results are equally relevant to the Le~Roy formalism.

Since the Le Roy formalism allows for more straightforward algebra when dealing with isotopomer shifts, the remainder of this work follows the Le Roy formalism for clarity, although the results of this section can also be derived starting from Eqn.~\ref{eqn:dunhamfullold} with equal validity.

Using Eqn.~\ref{eqn:dunhamfull}, the transition energy between a lower state $\Lambda'', \nu'', J''$ and an upper state $\Lambda', \nu', J'$ (the transition will be hence identified as $\Lambda'\leftarrow \Lambda''$) in isotopomer $X^{\prime}$ can thus be expressed as:
\begin{multline}\label{eqn:transition1}
    \big( E^{\Lambda^{\prime},X^{\prime}} - E^{\Lambda^{\prime\prime},X^{\prime}} \big) = h \nu^{X^{\prime}}_{\Lambda^{\prime}\leftarrow\Lambda^{\prime\prime}} = 
    \displaybreak[0]\\
    \sum_{(k,l) \neq (0,0)}  \Big(  \frac{\mu_{X^{\prime\prime}}}{\mu_{X^\prime}} \Big)^{l+k/2} \Big(\nu^{\prime} + \frac{1}{2} \Big)^{k} \big[ J^{\prime}(J^{\prime}+1)\big]^{l} Y^{\Lambda^{\prime}X^{\prime\prime}}_{kl}
    \displaybreak[0]\\
    + \sum_{(k,l) \geq (0,0)}  \Big(  \frac{\mu_{X^{\prime\prime}}}{\mu_{X^\prime}} \Big)^{l+k/2} \Big(\nu^{\prime} + \frac{1}{2} \Big)^{k} \big[ J^{\prime}(J^{\prime}+1)\big]^{l} 
    \displaybreak[0]\\
    \times \Bigg( \frac{\Delta M^{X^\prime}_{A}}{M^{X^\prime}_{A}}\delta^{\Lambda^{\prime},A}_{kl} + \frac{\Delta M^{X^\prime}_{B}}{M^{X^\prime}_{B}}\delta^{\Lambda^{\prime},B}_{kl} + \delta \langle r^2 \rangle ^{X^\prime}_{A} f^{\Lambda^{\prime}, A}_{kl} 
    \displaybreak[0]\\
    + \delta \langle r^2 \rangle ^{X^\prime}_{B} f^{\Lambda^{\prime}, B}_{kl} \Bigg)
    \displaybreak[0]\\
    - \sum_{(k,l) \neq (0,0)}  \Big(  \frac{\mu_{X^{\prime\prime}}}{\mu_{X^\prime}} \Big)^{l+k/2} \Big(\nu^{\prime\prime} + \frac{1}{2} \Big)^{k} \big[ J^{\prime\prime}(J^{\prime\prime}+1)\big]^{l} Y^{\Lambda^{\prime\prime}X^{\prime\prime}}_{kl}
    \displaybreak[0]\\
    - \sum_{(k,l) \geq (0,0)}  \Big(  \frac{\mu_{X^{\prime\prime}}}{\mu_{X^\prime}} \Big)^{l+k/2} \Big(\nu^{\prime\prime} + \frac{1}{2} \Big)^{k} \big[ J^{\prime\prime}(J^{\prime\prime}+1)\big]^{l} 
    \displaybreak[0]\\
    \times \Bigg( \frac{\Delta M^{X^\prime}_{A}}{M^{X^\prime}_{A}}\delta^{\Lambda^{\prime\prime},A}_{kl} + \frac{\Delta M^{X^\prime}_{B}}{M^{X^\prime}_{B}}\delta^{\Lambda^{\prime\prime},B}_{kl} + \delta \langle r^2 \rangle ^{X^\prime}_{A} f^{\Lambda^{\prime\prime}, A}_{kl} 
    \displaybreak[0]\\
    + \delta \langle r^2 \rangle ^{X^\prime}_{B} f^{\Lambda^{\prime\prime}, B}_{kl} \Bigg)
\end{multline}

In a spectroscopic experiment, which measures transition frequencies between molecular states, the highest order of correction in Eqn.~\ref{eqn:transition1} (that is, the maximum $k$ and $l$ in the summation series) is defined by the spectroscopic precision. Thus, the maximum $(k,l)$ is common to all summations in Eqn.~\ref{eqn:transition1}.

Based on Eqn.~\ref{eqn:transition1}, the isotope shift (of some isotopomer $X^{\prime}$ with respect to the reference isotopomer~$X^{\prime\prime}$, as per the definition) in the transition frequency between the two molecular states can be expressed as:

\begin{multline} \label{eqn:transition2}
    \big( E^{\Lambda^{\prime}, X^{\prime}} - E^{\Lambda^{\prime\prime}, X^{\prime}} \big) - \big( E^{\Lambda^{\prime}, X^{\prime\prime}} - E^{\Lambda^{\prime\prime}, X^{\prime\prime}} \big) = 
    \displaybreak[0]\\
    h \delta \nu^{X^{\prime} \leftarrow X^{\prime\prime}} _{\Lambda^{\prime}\leftarrow\Lambda^{\prime\prime}} = 
    \displaybreak[0]\\
    \sum_{(k,l) \neq (0,0)}  \Big(  \frac{\mu_{X^{\prime\prime}}}{\mu_{X^\prime}} \Big)^{l+k/2} \Big(\nu^{\prime} + \frac{1}{2} \Big)^{k} \big[ J^{\prime}(J^{\prime}+1)\big]^{l} Y^{\Lambda^{\prime}X^{\prime\prime}}_{kl}
    \displaybreak[0]\\
    + \sum_{(k,l) \geq (0,0)}  \Big(  \frac{\mu_{X^{\prime\prime}}}{\mu_{X^\prime}} \Big)^{l+k/2} \Big(\nu^{\prime} + \frac{1}{2} \Big)^{k} \big[ J^{\prime}(J^{\prime}+1)\big]^{l} 
    \displaybreak[0]\\
    \times \Bigg( \frac{\Delta M^{X^\prime}_{A}}{M^{X^\prime}_{A}}\delta^{\Lambda^{\prime},A}_{kl} + \frac{\Delta M^{X^\prime}_{B}}{M^{X^\prime}_{B}}\delta^{\Lambda^{\prime},B}_{kl} + \delta \langle r^2 \rangle ^{X^\prime}_{A} f^{\Lambda^{\prime}, A}_{kl} 
    \displaybreak[0]\\
    + \delta \langle r^2 \rangle ^{X^\prime}_{B} f^{\Lambda^{\prime}, B}_{kl} \Bigg)
    \displaybreak[0]\\
    - \sum_{(k,l) \neq (0,0)}  \Big(  \frac{\mu_{X^{\prime\prime}}}{\mu_{X^\prime}} \Big)^{l+k/2} \Big(\nu^{\prime\prime} + \frac{1}{2} \Big)^{k} \big[ J^{\prime\prime}(J^{\prime\prime}+1)\big]^{l} Y^{\Lambda^{\prime\prime}X^{\prime\prime}}_{kl}
    \displaybreak[0]\\
    - \sum_{(k,l) \geq (0,0)}  \Big(  \frac{\mu_{X^{\prime\prime}}}{\mu_{X^\prime}} \Big)^{l+k/2} \Big(\nu^{\prime\prime} + \frac{1}{2} \Big)^{k} \big[ J^{\prime\prime}(J^{\prime\prime}+1)\big]^{l} 
    \displaybreak[0]\\
    \times \Bigg( \frac{\Delta M^{X^\prime}_{A}}{M^{X^\prime}_{A}}\delta^{\Lambda^{\prime\prime},A}_{kl} + \frac{\Delta M^{X^\prime}_{B}}{M^{X^\prime}_{B}}\delta^{\Lambda^{\prime\prime},B}_{kl} + \delta \langle r^2 \rangle ^{X^\prime}_{A} f^{\Lambda^{\prime\prime}, A}_{kl} 
    \displaybreak[0]\\
    + \delta \langle r^2 \rangle ^{X^\prime}_{B} f^{\Lambda^{\prime\prime}, B}_{kl} \Bigg)
    \displaybreak[0]\\
    -\sum_{(k,l) \neq (0,0)}  \Big(  \frac{\mu_{X^{\prime\prime}}}{\mu_{X^{\prime\prime}}} \Big)^{l+k/2} \Big(\nu^{\prime} + \frac{1}{2} \Big)^{k} \big[ J^{\prime}(J^{\prime}+1)\big]^{l} Y^{\Lambda^{\prime}X^{\prime\prime}}_{kl}
    \displaybreak[0]\\
    - \sum_{(k,l) \geq (0,0)}  \Big(  \frac{\mu_{X^{\prime\prime}}}{\mu_{X^{\prime\prime}}} \Big)^{l+k/2} \Big(\nu^{\prime} + \frac{1}{2} \Big)^{k} \big[ J^{\prime}(J^{\prime}+1)\big]^{l} 
    \displaybreak[0]\\
    \times \Bigg( \frac{\Delta M^{X^{\prime\prime}}_{A}}{M^{X^{\prime\prime}}_{A}}\delta^{\Lambda^{\prime},A}_{kl} + \frac{\Delta M^{X^{\prime\prime}}_{B}}{M^{X^{\prime\prime}}_{B}}\delta^{\Lambda^{\prime},B}_{kl} + \delta \langle r^2 \rangle ^{X^{\prime\prime}}_{A} f^{\Lambda^{\prime}, A}_{kl} 
    \displaybreak[0]\\
    + \delta \langle r^2 \rangle ^{X^{\prime\prime}}_{B} f^{\Lambda^{\prime}, B}_{kl} \Bigg)
    \displaybreak[0]\\
    + \sum_{(k,l) \neq (0,0)}  \Big(  \frac{\mu_{X^{\prime\prime}}}{\mu_{X^{\prime\prime}}} \Big)^{l+k/2} \Big(\nu^{\prime\prime} + \frac{1}{2} \Big)^{k} \big[ J^{\prime\prime}(J^{\prime\prime}+1)\big]^{l} Y^{\Lambda^{\prime\prime}X^{\prime\prime}}_{kl}
    \displaybreak[0]\\
    + \sum_{(k,l) \geq (0,0)}  \Big(  \frac{\mu_{X^{\prime\prime}}}{\mu_{X^{\prime\prime}}} \Big)^{l+k/2} \Big(\nu^{\prime\prime} + \frac{1}{2} \Big)^{k} \big[ J^{\prime\prime}(J^{\prime\prime}+1)\big]^{l} 
    \displaybreak[0]\\
    \times \Bigg( \frac{\Delta M^{X^{\prime\prime}}_{A}}{M^{X^{\prime\prime}}_{A}}\delta^{\Lambda^{\prime\prime},A}_{kl} + \frac{\Delta M^{X^{\prime\prime}}_{B}}{M^{X^{\prime\prime}}_{B}}\delta^{\Lambda^{\prime\prime},B}_{kl} + \delta \langle r^2 \rangle ^{X^{\prime\prime}}_{A} f^{\Lambda^{\prime\prime}, A}_{kl} 
    \displaybreak[0]\\
    + \delta \langle r^2 \rangle ^{X^{\prime\prime}}_{B} f^{\Lambda^{\prime\prime}, B}_{kl} \Bigg)
\end{multline}

The energy terms in Eqn.~\ref{eqn:transition2} that correspond to the reference isotopomer can be simplified due to the definition of the relative terms. It is noticed that $\Big(\frac{\mu_{X^{\prime\prime}}}{\mu_{X^{\prime\prime}}} \Big)^{l+k/2} = 1$. Additionally, $\Delta M^{X^{\prime\prime}}_{A} = 0$ and $\Delta M^{X^{\prime\prime}}_{B} = 0$ by definition, and so the modified mass terms vanish from Eqn.~\ref{eqn:transition2}. Similarly, the changes in the mean-squared charge radii $\delta \langle r^2 \rangle ^{X^{\prime\prime}}_{A}$ and $\delta \langle r^2 \rangle ^{X^{\prime\prime}}_{A}$ for the reference isotopomer are by definition $0$. Therefore, Eqn.~\ref{eqn:transition2} becomes:
\begin{multline} \label{eqn:isotopeshift1}
    h \delta \nu^{X^{\prime} \leftarrow X^{\prime\prime}} _{\Lambda^{\prime}\leftarrow\Lambda^{\prime\prime}} = 
    \displaybreak[0]\\
    \sum_{(k,l) \neq (0,0)}  \Big(  \frac{\mu_{X^{\prime\prime}}}{\mu_{X^\prime}} \Big)^{l+k/2} \Big(\nu^{\prime} + \frac{1}{2} \Big)^{k} \big[ J^{\prime}(J^{\prime}+1)\big]^{l} Y^{\Lambda^{\prime}X^{\prime\prime}}_{kl}
    \displaybreak[0]\\
    + \sum_{(k,l) \geq (0,0)}  \Big(  \frac{\mu_{X^{\prime\prime}}}{\mu_{X^\prime}} \Big)^{l+k/2} \Big(\nu^{\prime} + \frac{1}{2} \Big)^{k} \big[ J^{\prime}(J^{\prime}+1)\big]^{l} 
    \displaybreak[0]\\
    \times \Bigg( \frac{\Delta M^{X^\prime}_{A}}{M^{X^\prime}_{A}}\delta^{\Lambda^{\prime},A}_{kl} + \frac{\Delta M^{X^\prime}_{B}}{M^{X^\prime}_{B}}\delta^{\Lambda^{\prime},B}_{kl} + \delta \langle r^2 \rangle ^{X^\prime}_{A} f^{\Lambda^{\prime}, A}_{kl} 
    \displaybreak[0]\\
    + \delta \langle r^2 \rangle ^{X^\prime}_{B} f^{\Lambda^{\prime}, B}_{kl} \Bigg)
    \displaybreak[0]\\
    - \sum_{(k,l) \neq (0,0)}  \Big(  \frac{\mu_{X^{\prime\prime}}}{\mu_{X^\prime}} \Big)^{l+k/2} \Big(\nu^{\prime\prime} + \frac{1}{2} \Big)^{k} \big[ J^{\prime\prime}(J^{\prime\prime}+1)\big]^{l} Y^{\Lambda^{\prime\prime}X^{\prime\prime}}_{kl}
    \displaybreak[0]\\
    - \sum_{(k,l) \geq (0,0)}  \Big(  \frac{\mu_{X^{\prime\prime}}}{\mu_{X^\prime}} \Big)^{l+k/2} \Big(\nu^{\prime\prime} + \frac{1}{2} \Big)^{k} \big[ J^{\prime\prime}(J^{\prime\prime}+1)\big]^{l} 
    \displaybreak[0]\\
    \times \Bigg( \frac{\Delta M^{X^\prime}_{A}}{M^{X^\prime}_{A}}\delta^{\Lambda^{\prime\prime},A}_{kl} + \frac{\Delta M^{X^\prime}_{B}}{M^{X^\prime}_{B}}\delta^{\Lambda^{\prime\prime},B}_{kl} + \delta \langle r^2 \rangle ^{X^\prime}_{A} f^{\Lambda^{\prime\prime}, A}_{kl} 
    \displaybreak[0]\\
    + \delta \langle r^2 \rangle ^{X^\prime}_{B} f^{\Lambda^{\prime\prime}, B}_{kl} \Bigg)
    \displaybreak[0]\\
    -\sum_{(k,l) \neq (0,0)}  \Big(\nu^{\prime} + \frac{1}{2} \Big)^{k} \big[ J^{\prime}(J^{\prime}+1)\big]^{l} Y^{\Lambda^{\prime}X^{\prime\prime}}_{kl}
    \displaybreak[0]\\
    + \sum_{(k,l) \neq (0,0)} \Big(\nu^{\prime\prime} + \frac{1}{2} \Big)^{k} \big[ J^{\prime\prime}(J^{\prime\prime}+1)\big]^{l} Y^{\Lambda^{\prime\prime}X^{\prime\prime}}_{kl}
\end{multline}

At this stage, the approximation that the reduced masses of the two isotopomers are equal needs to be introduced:~$\mu_{X^{\prime}}=\mu_{X^{\prime\prime}}$. As a result of this approximation, it follows that~$\Big(  \frac{\mu_{X^{\prime\prime}}}{\mu_{X^\prime}} \Big)^{l+k/2} = 1$ and Eqn.~\ref{eqn:isotopeshift1} becomes:
\begin{multline} \label{eqn:isotopeshift2}
    h \delta \nu^{X^{\prime} \leftarrow X^{\prime\prime}} _{\Lambda^{\prime}\leftarrow\Lambda^{\prime\prime}} = 
    \displaybreak[0]\\
    \sum_{(k,l) \neq (0,0)} \Big(\nu^{\prime} + \frac{1}{2} \Big)^{k} \big[ J^{\prime}(J^{\prime}+1)\big]^{l} Y^{\Lambda^{\prime}X^{\prime\prime}}_{kl}
    \displaybreak[0]\\
    + \sum_{(k,l) \geq (0,0)} \Big(\nu^{\prime} + \frac{1}{2} \Big)^{k} \big[ J^{\prime}(J^{\prime}+1)\big]^{l} 
    \displaybreak[0]\\
    \times \Bigg( \frac{\Delta M^{X^\prime}_{A}}{M^{X^\prime}_{A}}\delta^{\Lambda^{\prime},A}_{kl} + \frac{\Delta M^{X^\prime}_{B}}{M^{X^\prime}_{B}}\delta^{\Lambda^{\prime},B}_{kl} + \delta \langle r^2 \rangle ^{X^\prime}_{A} f^{\Lambda^{\prime}, A}_{kl} 
    \displaybreak[0]\\
    + \delta \langle r^2 \rangle ^{X^\prime}_{B} f^{\Lambda^{\prime}, B}_{kl} \Bigg)
    \displaybreak[0]\\
    - \sum_{(k,l) \neq (0,0)} \Big(\nu^{\prime\prime} + \frac{1}{2} \Big)^{k} \big[ J^{\prime\prime}(J^{\prime\prime}+1)\big]^{l} Y^{\Lambda^{\prime\prime}X^{\prime\prime}}_{kl}
    \displaybreak[0]\\
    - \sum_{(k,l) \geq (0,0)} \Big(\nu^{\prime\prime} + \frac{1}{2} \Big)^{k} \big[ J^{\prime\prime}(J^{\prime\prime}+1)\big]^{l} 
    \displaybreak[0]\\
    \times \Bigg( \frac{\Delta M^{X^\prime}_{A}}{M^{X^\prime}_{A}}\delta^{\Lambda^{\prime\prime},A}_{kl} + \frac{\Delta M^{X^\prime}_{B}}{M^{X^\prime}_{B}}\delta^{\Lambda^{\prime\prime},B}_{kl} + \delta \langle r^2 \rangle ^{X^\prime}_{A} f^{\Lambda^{\prime\prime}, A}_{kl} 
    \displaybreak[0]\\
    + \delta \langle r^2 \rangle ^{X^\prime}_{B} f^{\Lambda^{\prime\prime}, B}_{kl} \Bigg)
    \displaybreak[0]\\
    -\sum_{(k,l) \neq (0,0)}  \Big(\nu^{\prime} + \frac{1}{2} \Big)^{k} \big[ J^{\prime}(J^{\prime}+1)\big]^{l} Y^{\Lambda^{\prime}X^{\prime\prime}}_{kl}
    \displaybreak[0]\\
    + \sum_{(k,l) \neq (0,0)} \Big(\nu^{\prime\prime} + \frac{1}{2} \Big)^{k} \big[ J^{\prime\prime}(J^{\prime\prime}+1)\big]^{l} Y^{\Lambda^{\prime\prime}X^{\prime\prime}}_{kl}
\end{multline}
and therefore, in Eqn.~\ref{eqn:isotopeshift2}, the terms containing the Dunham parameters of the reference isotopomer~$Y^{\Lambda X^{\prime\prime}}_{kl}$ cancel out, leading to:

\begin{multline} \label{eqn:isotopeshift3}
    h \delta \nu^{X^{\prime} \leftarrow X^{\prime\prime}} _{\Lambda^{\prime}\leftarrow\Lambda^{\prime\prime}} = 
    \displaybreak[0]\\
    \sum_{(k,l) \geq (0,0)} \Big(\nu^{\prime} + \frac{1}{2} \Big)^{k} \big[ J^{\prime}(J^{\prime}+1)\big]^{l} 
    \displaybreak[0]\\
    \times \Bigg( \frac{\Delta M^{X^\prime}_{A}}{M^{X^\prime}_{A}}\delta^{\Lambda^{\prime},A}_{kl} + \frac{\Delta M^{X^\prime}_{B}}{M^{X^\prime}_{B}}\delta^{\Lambda^{\prime},B}_{kl} + \delta \langle r^2 \rangle ^{X^\prime}_{A} f^{\Lambda^{\prime}, A}_{kl} 
    \displaybreak[0]\\
    + \delta \langle r^2 \rangle ^{X^\prime}_{B} f^{\Lambda^{\prime}, B}_{kl} \Bigg)
    \displaybreak[0]\\
    - \sum_{(k,l) \geq (0,0)} \Big(\nu^{\prime\prime} + \frac{1}{2} \Big)^{k} \big[ J^{\prime\prime}(J^{\prime\prime}+1)\big]^{l} 
    \displaybreak[0]\\
    \times \Bigg( \frac{\Delta M^{X^\prime}_{A}}{M^{X^\prime}_{A}}\delta^{\Lambda^{\prime\prime},A}_{kl} + \frac{\Delta M^{X^\prime}_{B}}{M^{X^\prime}_{B}}\delta^{\Lambda^{\prime\prime},B}_{kl} + \delta \langle r^2 \rangle ^{X^\prime}_{A} f^{\Lambda^{\prime\prime}, A}_{kl} 
    \displaybreak[0]\\
    + \delta \langle r^2 \rangle ^{X^\prime}_{B} f^{\Lambda^{\prime\prime}, B}_{kl} \Bigg)
\end{multline}

Without loss of generality, it can be considered that only one atom in the diatomic molecule undergoes isotopic substitution, while the other atom is kept at its reference isotope (a reference isotopomer that differs in just one of the two atoms can always be chosen). Therefore, considering atom~$A$ as undergoing isotopic substitution, then $\Delta M^{X^\prime}_{B}=0$ and similarly $\delta \langle r^2 \rangle ^{X^\prime}_{B}=0$. Therefore, the isotope shift in a diatomic molecule where only one atom undergoes isotopic substitution, can be expressed as:
\begin{multline} \label{eqn:isotopeshift_prefinal}
    h \delta \nu^{X^{\prime} \leftarrow X^{\prime\prime}} _{\Lambda^{\prime}\leftarrow\Lambda^{\prime\prime}} = 
    \displaybreak[0]\\
    \sum_{(k,l) \geq (0,0)} \Big(\nu^{\prime} + \frac{1}{2} \Big)^{k} \big[ J^{\prime}(J^{\prime}+1)\big]^{l} 
    \displaybreak[0]\\
    \times \Bigg( \frac{M^{\alpha^{\prime\prime}}_{A}-M^{\alpha^\prime}_{A}}{M^{\alpha^\prime}_{A}}\delta^{\Lambda^{\prime},A}_{kl} + \delta \langle r^2 \rangle ^{X^\prime}_{A} f^{\Lambda^{\prime}, A}_{kl} \Bigg)
    \displaybreak[0]\\
    - \sum_{(k,l) \geq (0,0)} \Big(\nu^{\prime\prime} + \frac{1}{2} \Big)^{k} \big[ J^{\prime\prime}(J^{\prime\prime}+1)\big]^{l} 
    \displaybreak[0]\\
    \times \Bigg( \frac{M^{\alpha^{\prime\prime}}_{A}-M^{\alpha^\prime}_{A}}{M^{\alpha^\prime}_{A}}\delta^{\Lambda^{\prime\prime},A}_{kl} + \delta \langle r^2 \rangle ^{X^\prime}_{A} f^{\Lambda^{\prime\prime}, A}_{kl} \Bigg)    
\end{multline}

With a simple rearrangement, the isotope shift for a molecular transition $\Lambda^{\prime} \leftarrow \Lambda^{\prime\prime}$ between isotopomers $X^{\prime}$ and $X^{\prime\prime}$ where atom~$A$ contains isotopes $\alpha^{\prime}$ and $\alpha^{\prime\prime}$, respectively, can be finally expressed in a linear form with respect to the change in mean-squared nuclear charge radius:
\begin{equation} \label{eqn:isotopeshift_final}
    \delta \nu^{X^{\prime} \leftarrow X^{\prime\prime}} _{\Lambda^{\prime}\leftarrow\Lambda^{\prime\prime}} = \Psi \delta \langle r^2 \rangle ^{\alpha^\prime, \alpha^{\prime\prime}}_{A} + \Xi \frac{M^{\alpha^{\prime\prime}}_{A}-M^{\alpha^\prime}_{A}}{M^{\alpha^{\prime\prime}}_{A}M^{\alpha^\prime}_{A}}
\end{equation}
where
\begin{multline*}
    \Psi = \frac{1}{h} \sum_{(k,l) \geq (0,0)} \Bigg\{ \Big(\nu^{\prime} + \frac{1}{2} \Big)^{k} \big[ J^{\prime}(J^{\prime}+1)\big]^{l}  f^{\Lambda^{\prime}, A}_{kl}
    \displaybreak[0]\\
    - \Big(\nu^{\prime\prime} + \frac{1}{2} \Big)^{k} \big[ J^{\prime\prime}(J^{\prime\prime}+1)\big]^{l} f^{\Lambda^{\prime\prime}, A}_{kl}  \Bigg\}
\end{multline*}
and
\begin{multline*}
    \Xi = \frac{M^{\alpha^{\prime\prime}}_{A}}{h} \sum_{(k,l) \geq (0,0)} \Bigg\{ \Big(\nu^{\prime} + \frac{1}{2} \Big)^{k} \big[ J^{\prime}(J^{\prime}+1)\big]^{l}  \delta^{\Lambda^{\prime}, A}_{kl}
    \displaybreak[0]\\
    - \Big(\nu^{\prime\prime} + \frac{1}{2} \Big)^{k} \big[ J^{\prime\prime}(J^{\prime\prime}+1)\big]^{l} \delta^{\Lambda^{\prime\prime}, A}_{kl}  \Bigg\}
\end{multline*}
are isotope-independent constants. For the linearized isotope-shift equation in diatomic molecules (Eqn.~\ref{eqn:isotopeshift_final}), $\Psi$ is the field-shift factor and $\Xi$ is the mass-shift factor, analogous to the atomic counterparts. The definition of $\Xi$ contains the atomic mass of the reference isotope of atom~$A$; the term was added such that Eqn.~\ref{eqn:isotopeshift_final} is expressed in terms of $\frac{M^{\alpha^{\prime\prime}}_{A}-M^{\alpha^\prime}_{A}}{M^{\alpha^{\prime\prime}}_{A}M^{\alpha^\prime}_{A}}$, as in the atomic case (Eqn.~\ref{eqn:atomic_is}). Therefore, when a change of reference isotopomer is performed, the mass-shift factor should be scaled accordingly for the new reference mass.

Combining Eqns.~\ref{eqn:atomic_is} and~\ref{eqn:isotopeshift_final}, a King-plot analysis using atomic and molecular isotope shifts can be performed in the same manner as the analysis containing only atomic transitions, ultimately arriving at an expression identical to Eqn.~\ref{eqn:kingplot_atom} that relates the atomic and molecular isotope-shift factors.

\begin{figure}[h]
    \centering
    \includegraphics[width=8.6cm]{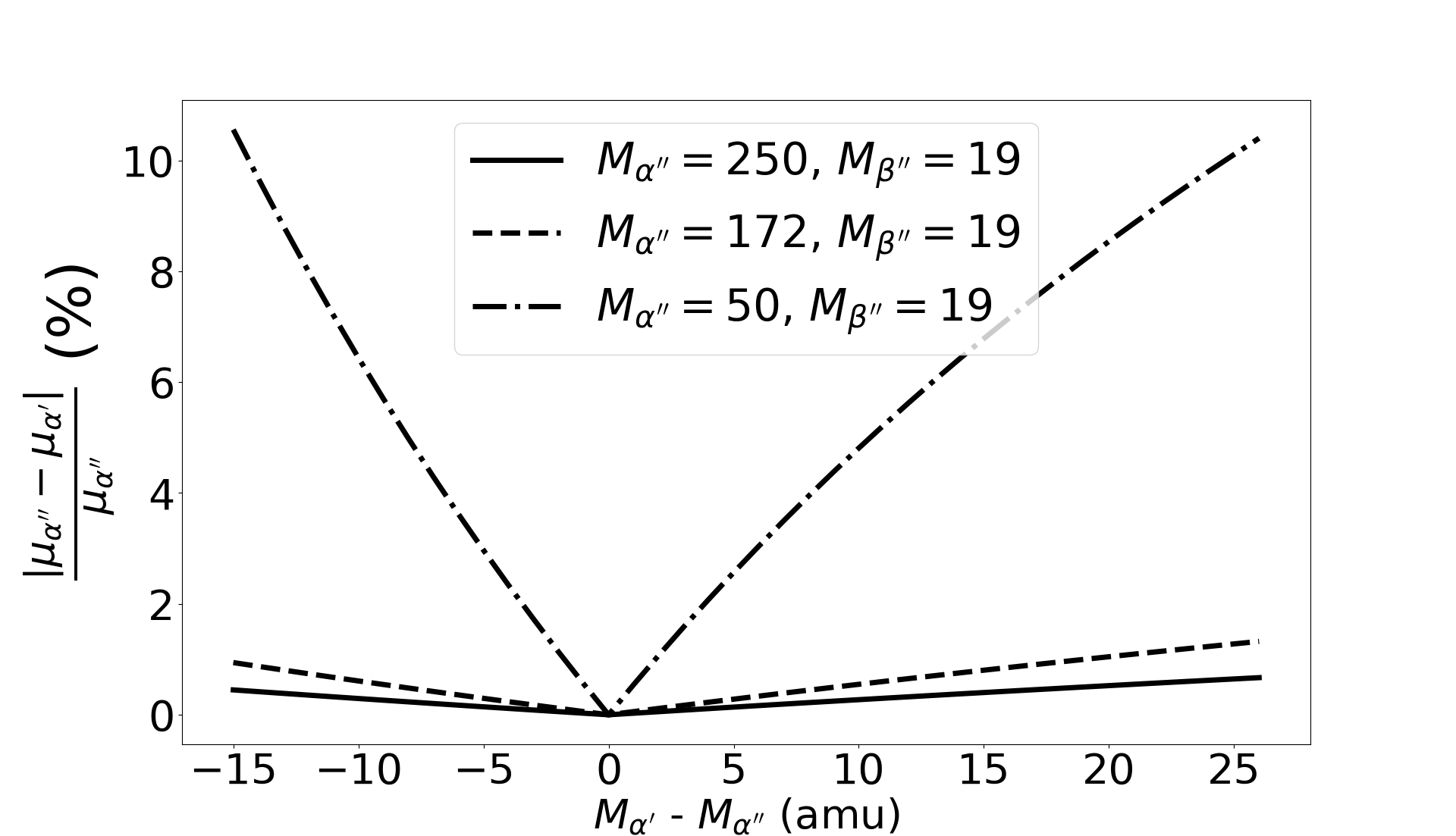}
    \caption{\small Percentage residual error for the approximation of constant reduced mass for molecules with different atomic-mass combinations in the medium-mass and heavy regions.}
    \label{fig:new_mu_approx_1}
\end{figure}

\begin{figure}[h]
    \centering
    \includegraphics[width=8.6cm]{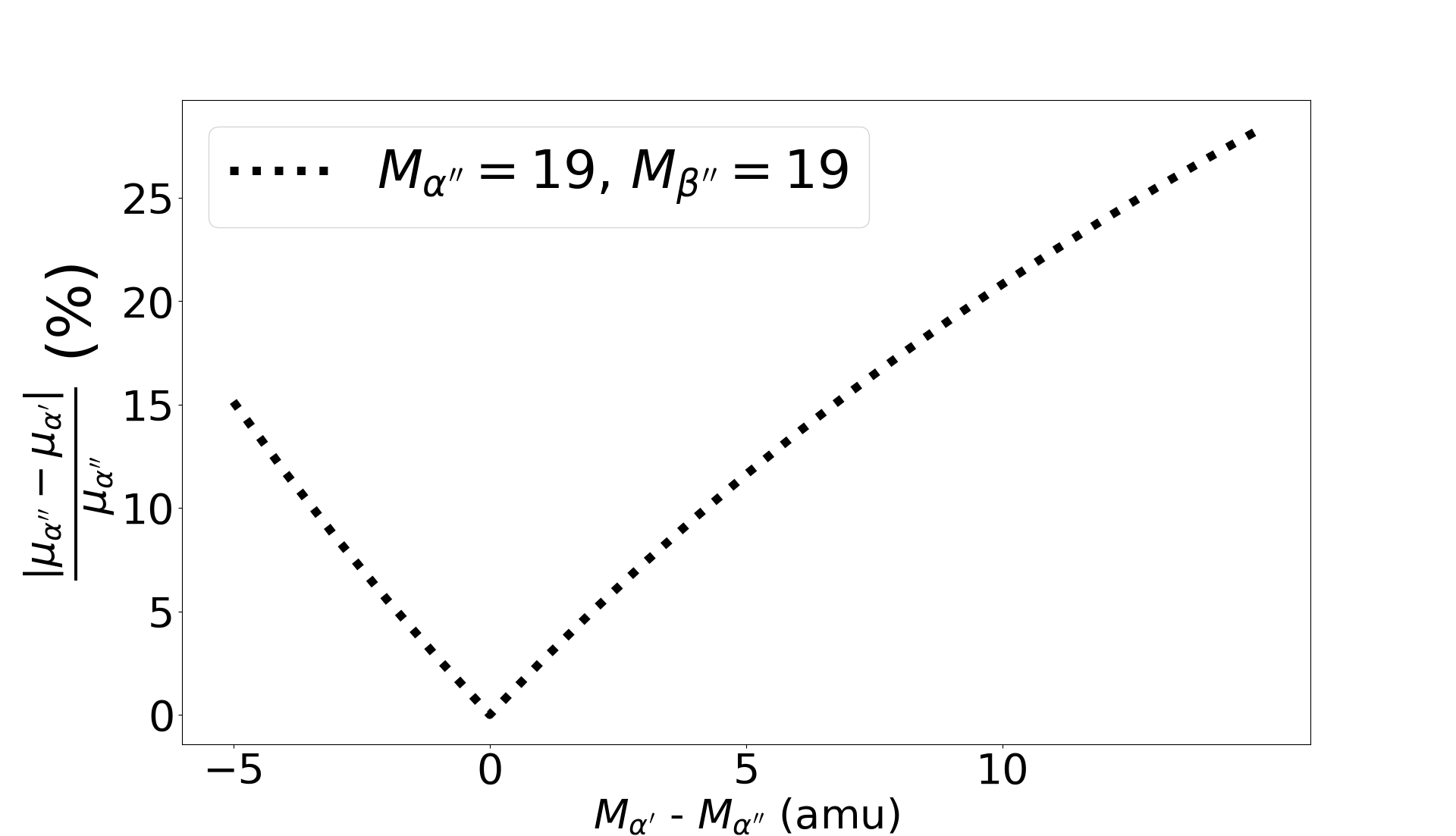}
    \caption{\small Percentage residual error for the approximation of the constant reduced mass for a representative light molecule.}
    \label{fig:new_mu_approx_2}
\end{figure}

The approximation~$\mu_{X^{\prime}}=\mu_{X^{\prime\prime}}$ that was used to arrive at Eqn.~\ref{eqn:isotopeshift_final} will introduce an error to the extracted $\Psi$ and $\Xi$ proportional to $\mu_{X^{\prime}} - \mu_{X^{\prime\prime}}$; that is, the deviation of $\frac{\mu_{X^{\prime\prime}}}{\mu_{X^{\prime}}}$ from unity. In Figs.~\ref{fig:new_mu_approx_1} and \ref{fig:new_mu_approx_2}, the percentage residual error of the approximation is shown for different atomic masses in diatomic molecules. For medium-mass and heavy diatomic molecules, the residual error remains below 10$\%$ even up to 20~amu from the reference isotope, which is lower or comparable to the typical error in atomic isotope-shift factors \cite{Cheal2012, Ohayon2022}.
% Since the error scales more rapidly when the substituted atom is lighter than its reference isotope, it appears wise to choose a reference on the lighter side of the available isotopes when analyzing molecular isotope shifts.

For light molecules, the error quickly rises above 10$\%$. While such a level can be comparable to the uncertainty in calculated atomic isotope-shift factors for complex atoms (especially for the specific mass shift)~\cite{Cheal2012}, it is recommended that the analysis of light molecules with the King-plot method is segmented. That is, the isotopic chain should be separated in smaller regions of only few amu each, and a separate King-plot analysis is performed in each, yielding a set of $\Psi$ and $\Xi$ for each segment of the isotopic chain.

%
%
%
%
%
% \newpage
\section{\label{sec:analysis}Analysis of molecular isotope shifts}
\subsection{\label{subsec:ybf}Optical isotope shifts in YbF}
To demonstrate the validity of Eqn.~\ref{eqn:isotopeshift_final}, literature high-resolution laser-spectroscopic data of the optical isotope shifts in $^{170-174,176}$YbF~\cite{Steimle2007} were analyzed with the molecular King-plot method along with literature high-resolution atomic measurements in Yb$^{+}$\cite{martenssonpendrill1994}. YbF possesses a simple electronic structure, with a single unpaired electron outside a closed shell, similar to the structure of Group II monofluorides~\cite{Sauer1996}, while the Yb element possesses multiple stable isotopes. Therefore, YbF is a suitable first test case to explore the validity of Eqn.~\ref{eqn:isotopeshift_final} in a diatomic molecule.

\begin{table*}[]
\renewcommand{\arraystretch}{1.65}
\parbox{\textwidth}{\caption{Parameter values (in cm$^{-1}$) and isotope shifts (in GHz) for the 369.4-nm transition in Yb$^{+}$~\cite{martenssonpendrill1994}, and the $T_0$ fitted parameter~\cite{Steimle2007Erratum}, the $^O P_{12}$(3)(2$\rightarrow$2)$_{A,0 \leftarrow X,0}$, and the $^O P_{12}$(9)(8$\rightarrow$9)$_{A,0 \leftarrow X,0}$ branch features in YbF~\cite{Steimle2007}.}}
 \label{tbl:sourcedata}
\begin{tabular}{ccccccccrr}
\hline \hline
\multicolumn{1}{c}{A} & \multicolumn{1}{c}{$T_0$} & \multicolumn{1}{c}{$^O P_{12}$(3)} & \multicolumn{1}{c}{$^O P_{12}$(9)} & \multicolumn{1}{c}{\footnote{ There is a sign change in the isotope shifts in Yb$^{+}$ with respect to Ref.~\cite{martenssonpendrill1994}, due to a difference in the subtraction convention.}$\delta \nu ^{A,172}_{\mathrm{Yb}^{+},369.4}$} & \multicolumn{1}{c}{$\delta \nu ^{A, 172}_{T_0}$} & \multicolumn{1}{c}{$\delta \nu ^{A, 172} _ {^{O}P_{12}(3)}$} & \multicolumn{1}{l}{$\delta \nu ^{A, 172} _ {^{O}P_{12}(9)}$} & \multicolumn{1}{l}{}  \\ [0.3cm] \hline
170                    & 18788.6634(39)                       & 18103.9508(15)  & 18098.7484(15)                                & +1.6223(8)                                                & +1.058(121)                                              & +1.019(77)           & +0.878(77)                                                                            &                      &                      \\ [0.15cm] \cline{1-8}
171                    & 18788.6502(4)                        &            &                                     & +1.0343(8)                                                & +0.663(32)                                               &                                                                                 &                      &                      &                      \\ [0.15cm] \cline{1-8}
172                    & 18788.6281(10)                       & 18103.9168(21)   & 18098.7191(21)                               & 0(0)                                                     & 0(0)                                                    & 0(0)               &   0(0)                                                                            &                      &                      \\ [0.15cm] \cline{1-8}
173                    & 18788.6145(7)                         &                 &                                & -0.5699(7)                                               & -0.408(37)                                              &                                                                                     &                      &                      &                      \\ [0.15cm] \cline{1-8}
174                    & 18788.5985(11)                       & 18103.8891(23)   & 18098.6982(23)                               & -1.2753(7)                                               & -0.887(45)                                              & -0.830(93)                & -0.627(93)                                                                          &                      &                      \\ [0.15cm] \cline{1-8}
176                    & 18788.5702(10)                       & 18103.8626(22)  & 18098.6779(22)                                & -2.4928(10)                                              & -1.736(42)                                              & -1.625(91)               & -1.235(91)                                                                  &                      &                     
\\
\hline \hline
\end{tabular}
\end{table*}

The YbF data were taken from Ref.~\cite{Steimle2007} and Table~1 of its Supplementary Material. King-plot analyses were performed using the fitted values of $T_{0}$ for the $A ^2 \Pi_{1/2}$~$(\nu=0)$ state in $^{170-174,176}$YbF~\cite{Steimle2007Erratum}, as well as the branch feature $^O P_{12}$(3) ($F^\prime=2,$ $F^{\prime\prime} = 2$) and the branch feature $^O P_{12}$(9) ($F^\prime=8,$ $F^{\prime\prime} = 9$) in the $A ^2 \Pi_{1/2}$~$(\nu=0) \leftarrow X ^2 \Sigma^{+}$~$(\nu=0)$ transition in $^{170,172,174,176}$YbF. 
%The source data, including the conversion to isotope shifts in units of GHz, is shown in Table~\ref{tbl:sourcedata}.

\begin{figure}[h]
    \centering
    \includegraphics[width=8.6cm]{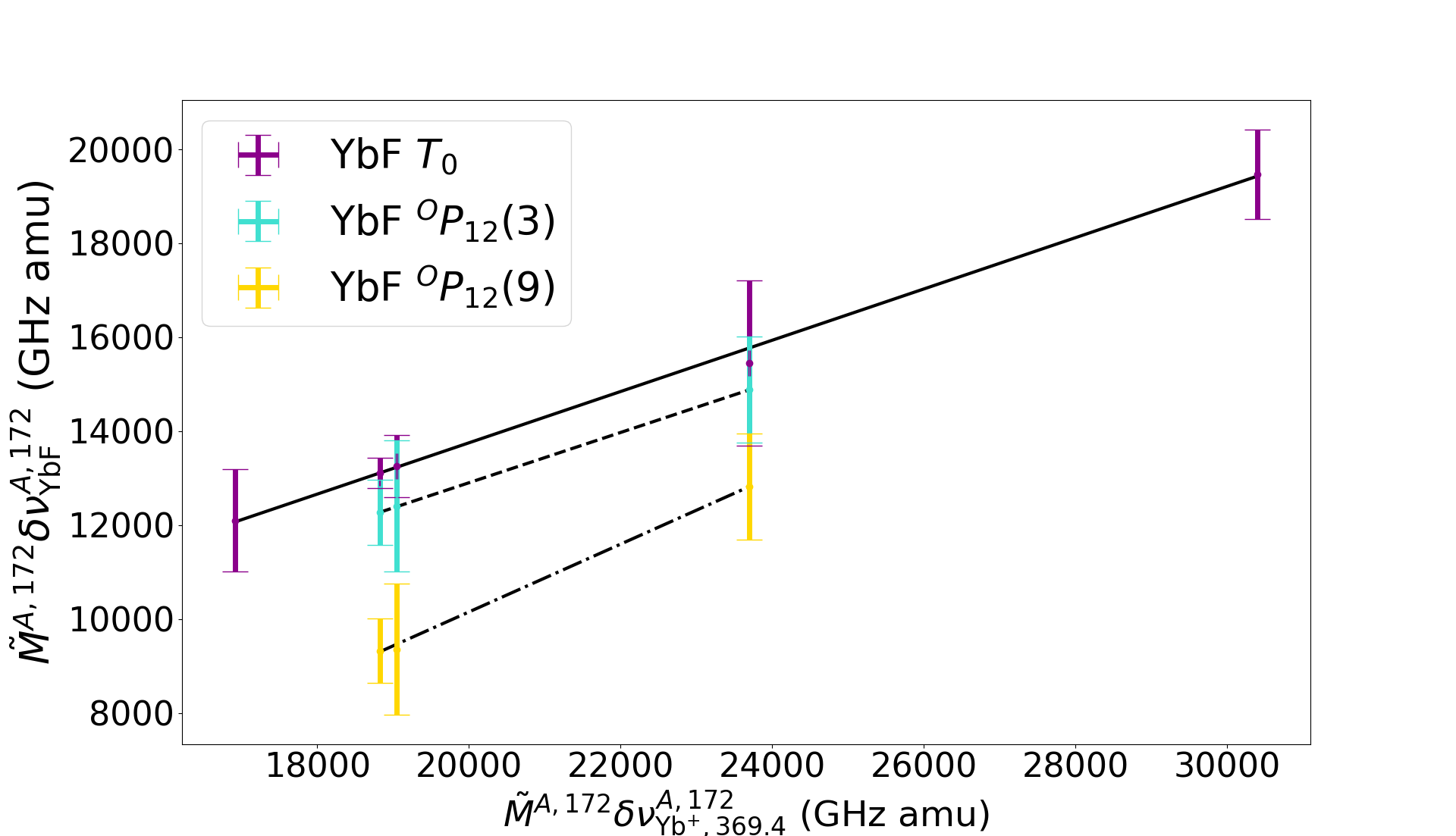}
    \caption{\small King plots for the isotope shifts in the $A ^2 \Pi_{1/2}$~$T_0$ values and the $^O P_{12}$(3)~($F^\prime=2,$ $F^{\prime\prime} = 2$)$_{A,0\leftarrow X,0}$ and $^O P_{12}$(9)~($F^\prime=8,$ $F^{\prime\prime} = 9$)$_{A,0\leftarrow X,0}$ branch features in YbF, and the 369.4-nm transition in Yb$^{+}$. All error bars arise from the 1$\sigma$ uncertainties of the isotope shifts and the uncertainties of the mass measurements taken from the 2020 Atomic Mass Evaluation~\cite{AME2020}.} %$\tilde{M}^{A,172}=\frac{M_{172}M_{A}}{M_{172}-M_{A}}$.}
    \label{fig:yb_ybf_king_plots}
\end{figure}

Using the mass values listed in the 2020 Atomic Mass Evaluation~\cite{AME2020}, three King plots were constructed for the YbF and Yb$^{+}$ transitions (Fig.~\ref{fig:yb_ybf_king_plots}). Evidently, the best-fit lines, obtained using an orthogonal-distance regression~(ODR) routine, pass through the error bars for all data points, demonstrating the linearity of the isotope-shift expression both in the molecule and in the ion.

As per Eqn.~\ref{eqn:kingplot_atom}, using which the field- and mass-shift of the molecule (Eqn.~\ref{eqn:isotopeshift_final}) are related to those of the ionic system, the slope~$m$ of the best-fit line~$y=mx+b$ in the King plot is equal to $\frac{F_i}{F_j}$ and the y-intercept to $b=K_i -\frac{F_i}{F_j}K_j$. The slopes and intercepts of the best-fit lines in Fig.~\ref{fig:yb_ybf_king_plots} are:
\begin{eqnarray}
    \nonumber
    m_{T_0} = +0.546(9) \\
    \nonumber
    b_{T_0} = +2830(180)
\end{eqnarray}
\begin{eqnarray}
    \nonumber
    m_{P_{12}(3)} = +0.535(2) \\
    \nonumber
   b_{P_{12}(3)} = +2200(50)
\end{eqnarray}
and
\begin{eqnarray}
    \nonumber
    m_{P_{12}(9)} = +0.722(22) \\
    \nonumber
   b_{P_{12}(9)} = -4300(400)
\end{eqnarray}
including the 1$\sigma$ errors from the ODR fit.

The field- and mass-shift parameters for the 369.4-nm transition in Yb$^{+}$ are known from Ref.~\cite{martenssonpendrill1994} (with adjusted sign for the reference convention):
\begin{eqnarray}
    \nonumber
    F_{369.4} = -14.185(190) \text{ GHz fm}^{-2} \\
    \nonumber
    K_{369.4} = -890(445) \text{ GHz amu}
\end{eqnarray}
The deduced molecular $\Psi$ and $\Xi$ are given in Table~\ref{tbl:ybf_factors}.

\begin{table}[h]
\renewcommand{\arraystretch}{1.65}
\caption{Field- ($\Psi$, in GHz~fm$^{-2}$) and mass-shift ($\Xi$, in GHz~amu) factors for the isotope shifts in the fitted $A ^2 \Pi_{1/2}$~$(\nu=0)$~T$_0$ parameter and the $A ^2 \Pi_{1/2}$~$(\nu=0) \leftarrow X ^2 \Sigma^{+}$~$(\nu=0)$~$^O P_{12}$(3) ($F^\prime=2,$ $F^{\prime\prime} = 2$) and $A ^2 \Pi_{1/2}$~$(\nu=0) \leftarrow X ^2 \Sigma^{+}$~$(\nu=0)$~$^O P_{12}$(9) ($F^\prime=8,$ $F^{\prime\prime} = 9$) branch features in YbF, from the King-plot analyses with the isotope shifts in the 369.4-nm transition in Yb$^{+}$.}
\label{tbl:ybf_factors}
\begin{tabular}{lrrr}
\hline \hline
                            & \multicolumn{1}{c}{\hspace{10pt}$T_0^{\mathrm{YbF}}$} & \multicolumn{1}{c}{\hspace{10pt}$^{O}P_{12}(3)^{\mathrm{YbF}}$} & \multicolumn{1}{c}{\hspace{10pt}$^{O}P_{12}(9)^{\mathrm{YbF}}$} \\ \hline
\multicolumn{1}{l}{$\Psi$} & \hspace{10pt}-7.75(17)                         &\hspace{10pt} -7.59(11)                     & -10.24(34)      \\
\multicolumn{1}{l}{$\Xi$}  &\hspace{10pt} +2350(310)                       &\hspace{10pt} +1720(240)
 & -4900(600) \\
 \hline \hline
\end{tabular}
\end{table}

\begin{figure}[h]
    \centering
    \includegraphics[width=8.6cm]{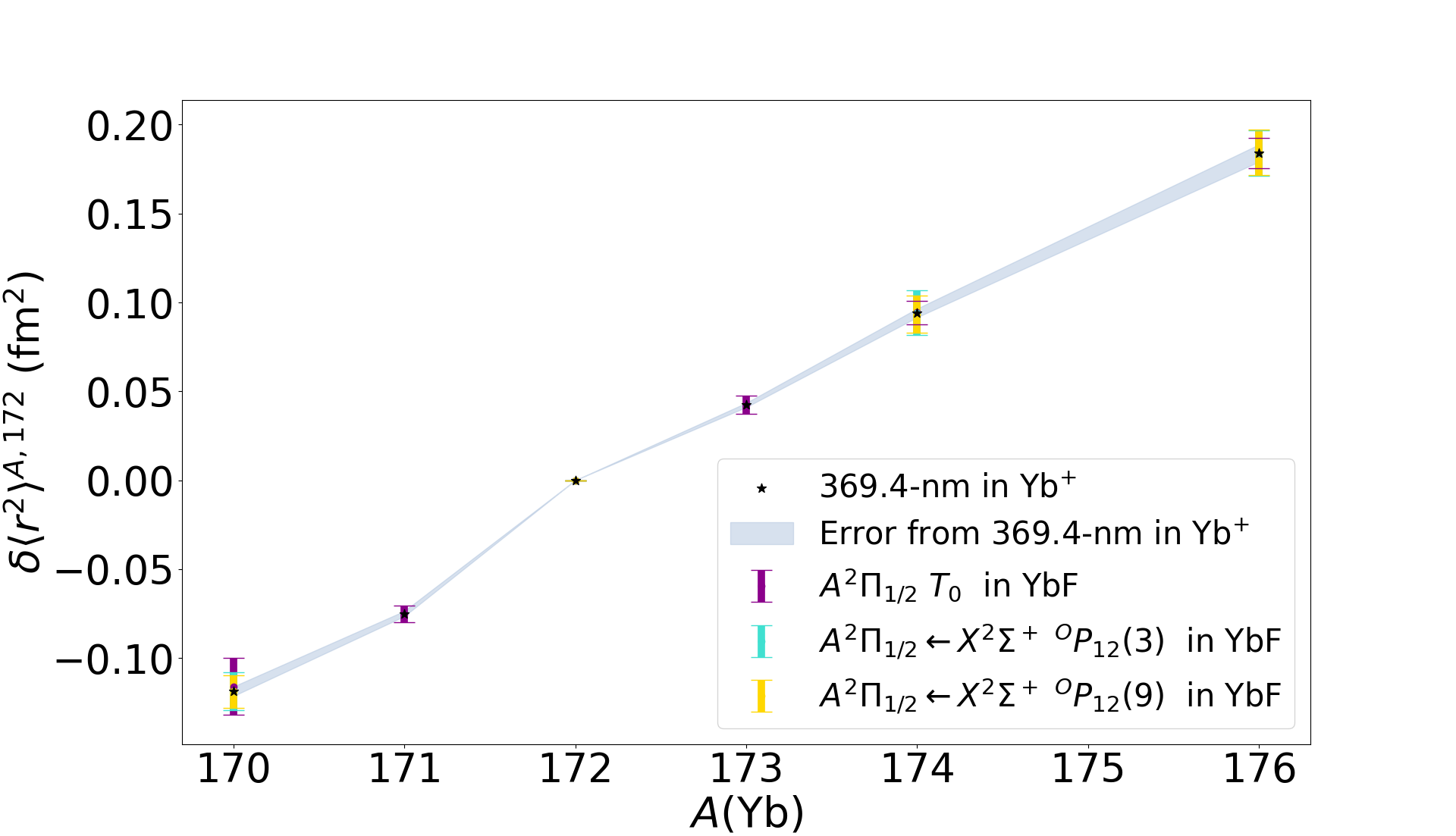}
    \caption{\small Changes in the mean-squared nuclear charge radii of $^{170-174,176}$Yb extracted from isotope shifts in YbF~\cite{Steimle2007} using the King-plot method (Eqn.~\ref{eqn:isotopeshift_final}), compared with those extracted from isotope-shift measurements in the 369.4-nm transition in Yb$^{+}$~\cite{martenssonpendrill1994}. The uncertainties (also given in Table~\ref{tbl:yb_radii}) stem from the experimental uncertainties in the isotope shifts and the molecular isotope shift factors extracted from the King plot (Table~\ref{tbl:ybf_factors}), using Equation~\ref{eqn:atomic_is}. The reference errors (grey band) are taken from Ref.~\cite{martenssonpendrill1994}.}
    \label{fig:yb_ybf_radii}
\end{figure}

Using the factors in Table~\ref{tbl:ybf_factors}, the changes in the mean-squared nuclear charge radii of the Yb isotopes can be extracted from the isotope shifts in YbF using Eqn.~\ref{eqn:isotopeshift_final}. The results, along with the values of $\delta \langle r^2 \rangle ^{A,172}$ extracted from the isotope shifts in the 369.4-nm transition in Yb$^{+}$ using $F_{369.4}$ and $K_{369.4}$ listed above, are presented in Table~\ref{tbl:yb_radii} and Fig.~\ref{fig:yb_ybf_radii}~\footnote{The mean-squared nuclear charge radii in Ref.~\cite{martenssonpendrill1994} are given with respect to $^{176}$Yb. There is a disagreement between the values of the mean-squared charge radii in Ref.~\cite{martenssonpendrill1994} and Refs.~\cite{Clark1979,Angeli2013}. An explanation for the discrepancy can be found in Ref.~\cite{martenssonpendrill1994}. The disagreement between the results of the two works falls beyond the scope of the present work, as the linearity of the molecular isotope shift would be valid for any choice of isotope-shift factors for Yb$^{+}$.}. 

\begin{table}[]
\renewcommand{\arraystretch}{1.65}
\caption{Changes in the mean-squared nuclear charge radii of Yb isotopes (in fm$^2$) extracted from the isotope shifts in the 369.4-nm transition in Yb$^{+}$~\cite{martenssonpendrill1994} and from the isotope shifts in YbF analyzed with the King-plot method.}
\label{tbl:yb_radii}
\begin{tabular}{ccccc}
\hline \hline
A                        & $\delta \langle r^2 \rangle ^{A,172} _{369.4}$ & $\delta \langle r^2 \rangle ^{A,172} _{P_{12}(3)}$ & $\delta \langle r^2 \rangle ^{A,172} _{P_{12}(9)}$ & $\delta \langle r^2 \rangle ^{A,172} _{T_{0}}$ \\ \hline
\multicolumn{1}{c}{170} & \multicolumn{1}{c}{-0.119(3)}                          & \multicolumn{1}{c}{-0.119(11)}    &         -0.119(9)             & -0.116(16)                                               \\
\multicolumn{1}{c}{171} & \multicolumn{1}{c}{-0.075(1)}                          & \multicolumn{1}{c}{}                 &                   & -0.075(5)                                                \\
\multicolumn{1}{c}{172} & \multicolumn{1}{c}{0(0)}                              & \multicolumn{1}{c}{0(0)}               &        0(0)         & 0(0)                                                     \\
\multicolumn{1}{c}{173} & \multicolumn{1}{c}{+0.042(1)}                           & \multicolumn{1}{c}{}              &                     & +0.043(5)                                                 \\
\multicolumn{1}{c}{174} & \multicolumn{1}{c}{+0.094(2)}                           & \multicolumn{1}{c}{+0.094(13)}       &       +0.093(10)             & +0.094(7)                                                 \\
\multicolumn{1}{c}{176} & \multicolumn{1}{c}{+0.184(5)}                           & \multicolumn{1}{c}{+0.184(13)}     &          +0.184(13)            & +0.184(9)                                               \\ \hline \hline
\end{tabular}
\end{table}

\begin{figure}[h]
    \centering
    \includegraphics[width=8.6cm]{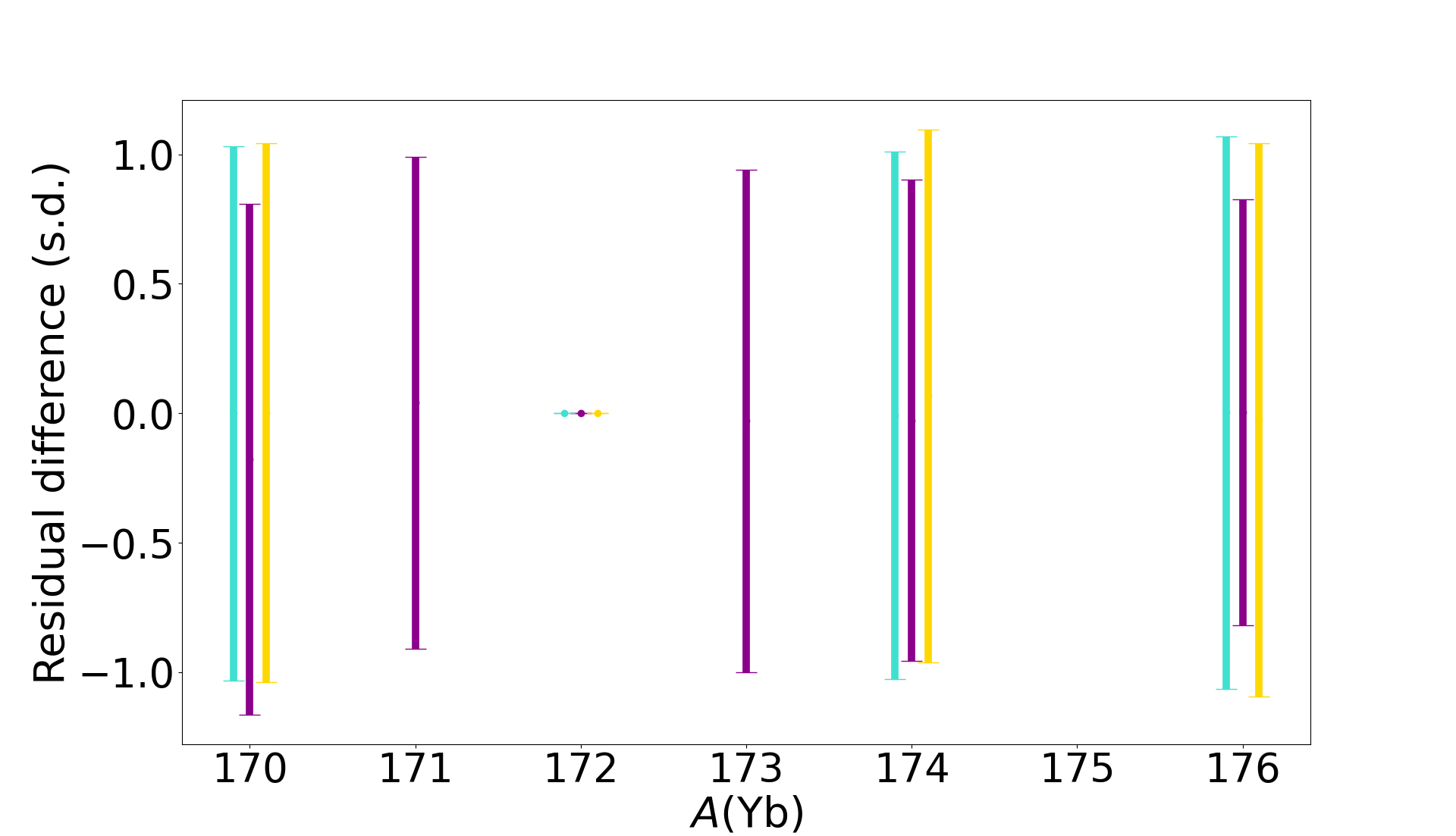}
    \caption{\small Residual difference of $\delta \langle r^2 \rangle_{\mathrm{Yb}} ^{A,172}$ extracted from the molecular isotope shifts compared with the values extracted from high resolution atomic spectroscopy~\cite{martenssonpendrill1994}, in units of the error in the molecular values. Magenta: $T_0$, cyan: $^O P_{12}$(3), gold: $^O P_{12}$(9). An arbitrary horizontal offset between the different sets of molecular isotope shifts has been applied for visual clarity. The error bars represent the uncertainties in the radii (Table~\ref{tbl:yb_radii}) extracted using the molecular isotope shifts, divided by the uncertainties in the radii using the atomic isotope shifts.}
    \label{fig:yb_ybf_radii_residuals}
\end{figure}

% In Fig.~\ref{fig:yb_ybf_radii}, the values of $\delta \langle r^2 \rangle_{\mathrm{Yb}} ^{A,172}$ extracted from the molecular isotope shifts using the King-plot method are compared with those extracted using the 369.4-nm transition in Yb$^{+}$ and the factors listed above. The residual difference in $\delta \langle r^2 \rangle_{\mathrm{Yb}} ^{A,172}$ between the values extracted from the molecular isotope shifts and those from the high-resolution spectroscopy of the 369.4-nm transition in Yb$^{+}$ is shown in Fig.~\ref{fig:yb_ybf_radii_residuals}, in units of the error of the molecular values. 
For all three sets of isotope shifts, the changes in mean-squared charge radii extracted from the molecule are consistent with the results from atomic spectroscopy within 1$\sigma$ of the molecular value. The residual errors are shown in Fig.~\ref{fig:yb_ybf_radii_residuals}.
% (taken from Ref.~\cite{martenssonpendrill1994}).

\subsection{\label{subsec:zro} Optical isotope shifts in ZrO}
To further explore the applicability of Eqn.~\ref{eqn:isotopeshift_final}, a King-plot analysis was also performed on literature optical isotopes shifts in the $P$(1), $R$(1), $P$(20), and $R$(20) rotational lines of the $C{}^{1}\Sigma^{+}-X{}^{1}\Sigma^{+}$ transition in ZrO~\cite{Simard1988} and the 327-nm transition in Zr$^{+}$~\cite{Campbell2002}. The molecular and atomic (ionic) isotope shifts for $^{90-92,94,96}$Zr are shown in Table~\ref{tbl:zr_isotope_shifts}. Simard et al.~\cite{Simard1988} did not report uncertainties for the observed wavenumbers appearing in Ref.~\cite{Simard1988}. For the purpose of this demonstrative analysis, an uncertainty of 50 MHz was associated with each isotope-shift value, since the present analysis aims to explore the applicability of the molecular King-plot method, rather than the rigorous extraction of charge radii.

\begin{figure}[h]
    \centering
    \includegraphics[width=8.6cm]{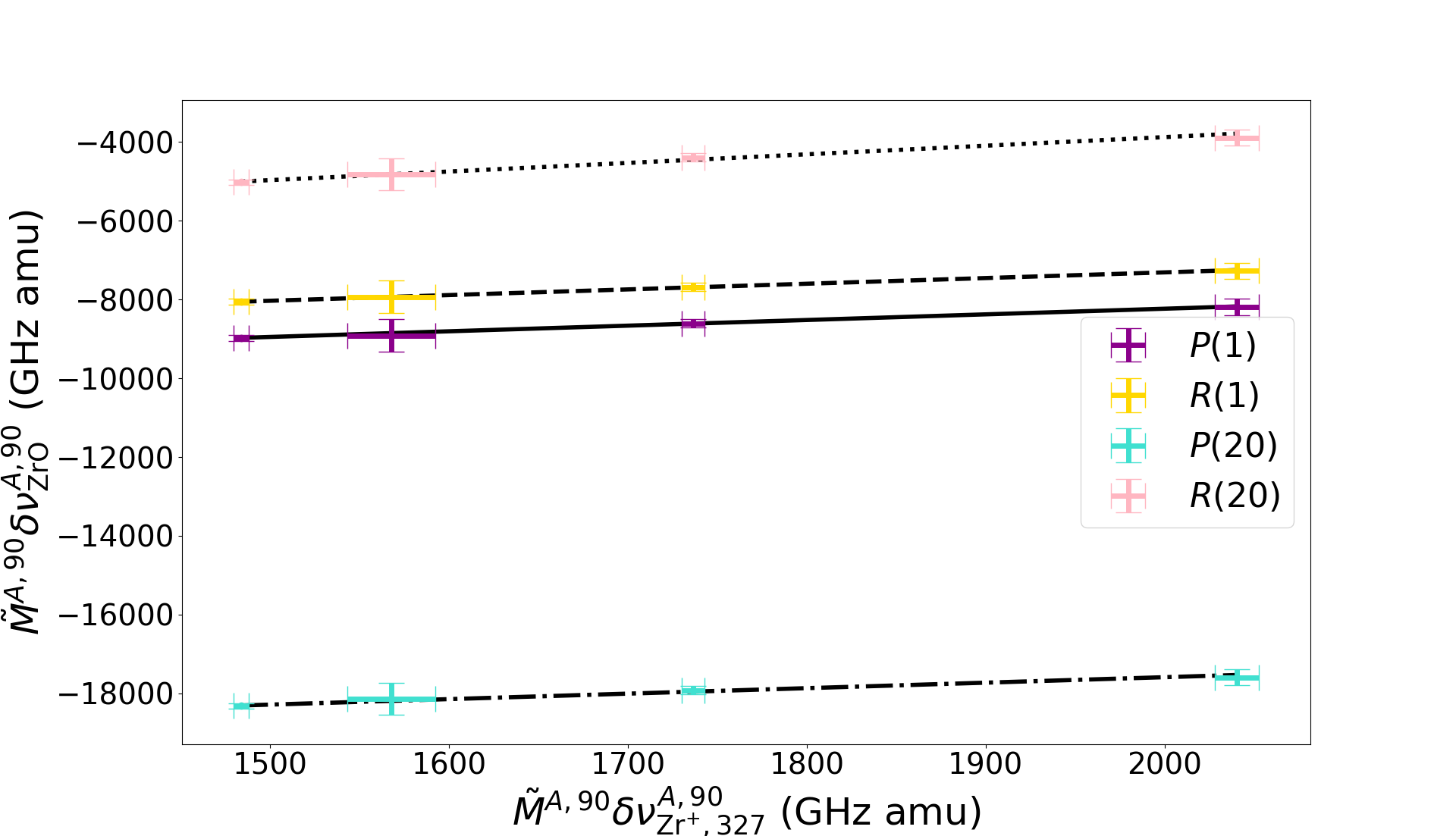}
    \caption{\small King plots for the $P$(1), $R$(1), $P$(20), and $R$(20) lines of the $C{}^{1}\Sigma^{+}-X{}^{1}\Sigma^{+}$ transition in ZrO~\cite{Simard1988} and the 327-nm transition in Zr$^{+}$~\cite{Campbell2002}. All error bars arise from the 1$\sigma$ uncertainties of the isotope shifts (Table~\ref{tbl:zr_isotope_shifts}) and the uncertainties of the mass measurements taken from the 2020 Atomic Mass Evaluation~\cite{AME2020}.}
    \label{fig:zr_zro_king_plots}
\end{figure}

\begin{figure}[h]
    \centering
    \includegraphics[width=8.6cm]{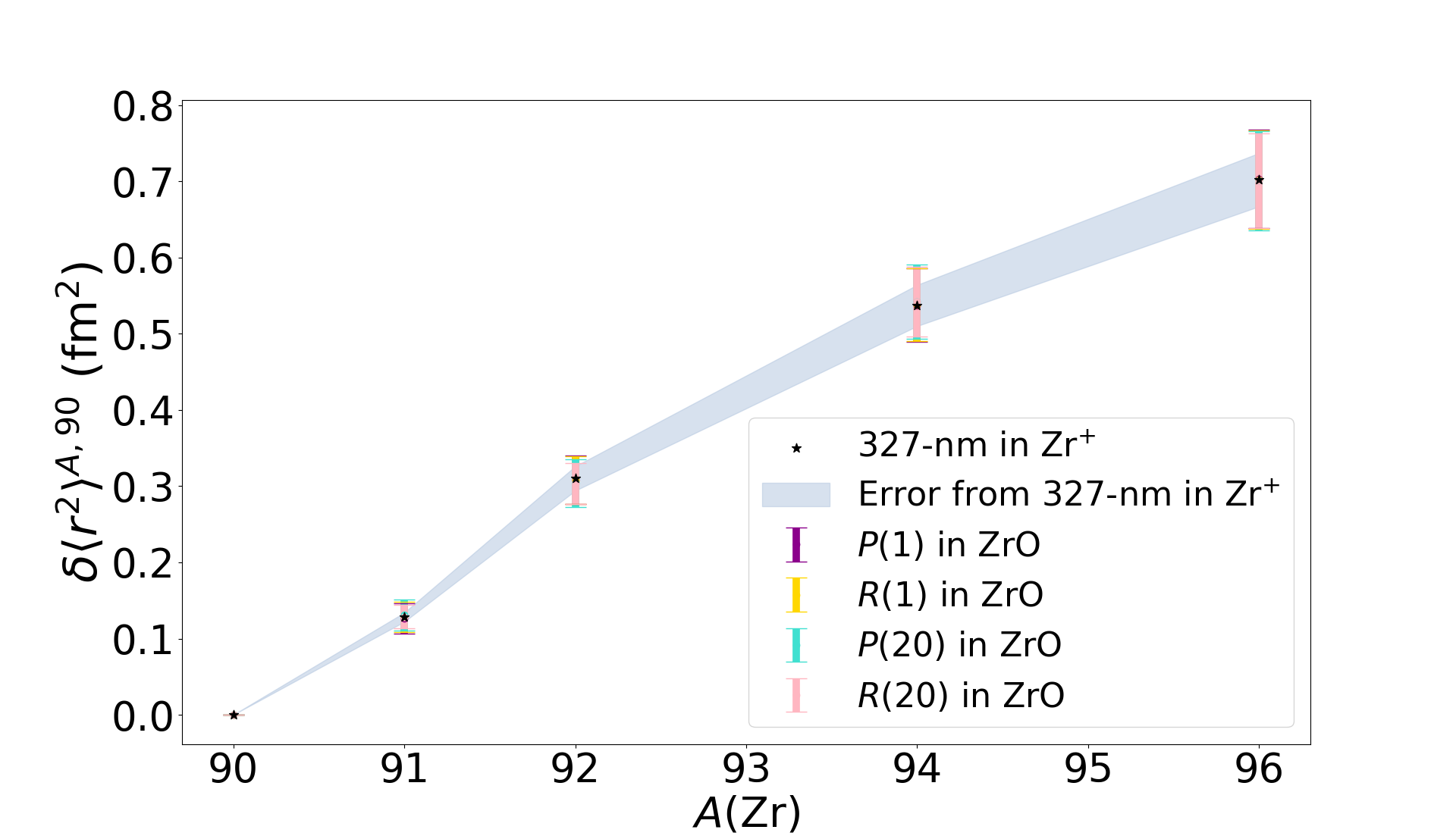}
    \caption{\small Comparison of the changes in the mean-squared nuclear charge radii $\delta \langle r^2 \rangle^{A,90}_\mathrm{Zr}$ of the stable Zr isotopes extracted using the 327-nm transition in Zr$^{+}$ and from four rotational lines in ZrO. The uncertainties (also given in Table~\ref{tbl:zr_radii}) stem from the experimental uncertainties in the isotope shifts (Table~\ref{tbl:zr_isotope_shifts}) and the molecular isotope shift factors extracted from the King plot (Table~\ref{tbl:zr_isotope_shift_factors}), using Equation 2. The reference errors (grey band) correspond to the uncertainty in the radii extracted from the 327-nm transition in Zr$^{+}$.}
    \label{fig:zr_zro_radii}
\end{figure}

The four King plots and a schematic comparison of $\delta \langle r^2 \rangle^{A,90}_\mathrm{Zr}$ are shown in Figs.~\ref{fig:zr_zro_king_plots} and \ref{fig:zr_zro_radii}, respectively. The isotope-shift factors $\Psi$ and $\Xi$ for the molecular transitions, extracted from the King-plot analysis, are shown in Table~\ref{tbl:zr_isotope_shift_factors}. As evident from Fig.~\ref{fig:zr_zro_radii} and Table~\ref{tbl:zr_radii}, the changes in mean-squared charge radii for $^{90-92, 94, 96}$Zr extracted from the four molecular transitions are consistent with the values extracted from the spectroscopy of Zr$^{+}$, falling within 1$\sigma$ of the values extracted from atomic spectroscopy.

\begin{table}[]
\renewcommand{\arraystretch}{1.65}
\caption{Isotope shifts $\delta \nu ^{A,90}$ (in GHz) for the 327-nm transition in Zr$^{+}$~\cite{Campbell2002} and the $P$(1), $R$(1), $P$(20), and $R$(20) rotational lines of the $C{}^{1}\Sigma^{+}-X{}^{1}\Sigma^{+}$ transition in ZrO~\cite{Simard1988} for $^{90-92,94,96}$Zr.}
\label{tbl:zr_isotope_shifts}
\begin{tabular}{cccccc}
\hline \hline
A & 327-nm & $P$(1)    & $R$(1)    & $P$(20)    & $R$(20)   \\ \hline
90      & 0(0)          & 0(0)      & 0(0)      & 0(0)       & 0(0)      \\
91      & -0.192(3)     & 1.091(50) & 0.971(50) & 2.221(50)  & 0.591(50) \\
92      & -0.494(3)     & 1.982(50) & 1.760(50) & 4.260(50)  & 0.941(50) \\
94      & -0.823(3)     & 4.077(50) & 3.640(50) & 8.496(50)  & 2.081(50) \\
96      & -1.033(3)     & 6.248(50) & 5.609(50) & 12.756(50) & 3.493(50) \\
\hline \hline
\end{tabular}
\end{table}

\begin{table}[]
\renewcommand{\arraystretch}{1.65}
\caption{Molecular isotope-shift factors $\Psi$ (in GHz fm$^{-2}$) and $\Xi$ (in GHz amu) for the four rotational transitions in ZrO analyzed with the King-plot method against the 327-nm isotope shifts in Zr$^{+}$.}
\label{tbl:zr_isotope_shift_factors}
\begin{tabular}{ccc}
\hline \hline
        & $\Psi$  & $\Xi$       \\ 
        \hline
$P$(1)  & -3.0(7) & -12000(800) \\
$R$(1)  & -3.0(7) & -11000(800) \\
$P$(20) & -2.9(7) & -21200(800) \\
$R$(20) & -4.6(8) & -9600(800) \\ \hline \hline
\end{tabular}
\end{table}

\begin{table}[]
\renewcommand{\arraystretch}{1.65}
\caption{Changes in the mean-squared nuclear charge radii of $^{90-92,94,96}$Zr extracted with laser spectroscopy using the 327-nm transition in Zr$^{+}$~\cite{Campbell2002} and four rotational lines in ZrO~\cite{Simard1988}.}
\label{tbl:zr_radii}
\begin{tabular}{cccccc}
\hline \hline
A  & $\delta \langle r^2 \rangle^{A,90}_{\mathrm{327}}$ & $\delta \langle r^2 \rangle^{A,90}_{P\mathrm{(1)}}$ & $\delta \langle r^2 \rangle^{A,90}_{R\mathrm{(1)}}$ & $\delta \langle r^2 \rangle^{A,90}_{P\mathrm{(20)}}$ & $\delta \langle r^2 \rangle^{A,90}_{R\mathrm{(20)}}$ \\  \hline
90 & 0(0)                                               & 0(0)                                                & 0(0)                                                & 0(0)                                                 & 0(0)                                                 \\
91 & 0.128(6)                                           & 0.126(20)                                           & 0.129(20)                                           & 0.131(21)                                            & 0.129(15)                                            \\
92 & 0.310(16)                                          & 0.308(32)                                           & 0.308(31)                                           & 0.304(31)                                            & 0.304(27)                                            \\
94 & 0.537(27)                                          & 0.537(48)                                           & 0.538(48)                                           & 0.541(49)                                            & 0.542(46)                                            \\
96 & 0.702(35)                                          & 0.703(65)                                           & 0.703(65)                                           & 0.701(65)                                            & 0.701(62)                       \\ \hline \hline                    
\end{tabular}
\end{table}

\subsection{\label{subsec:snh} Infrared isotope shifts in SnH}

\begin{figure}[h]
    \centering
    \includegraphics[width=8.6cm]{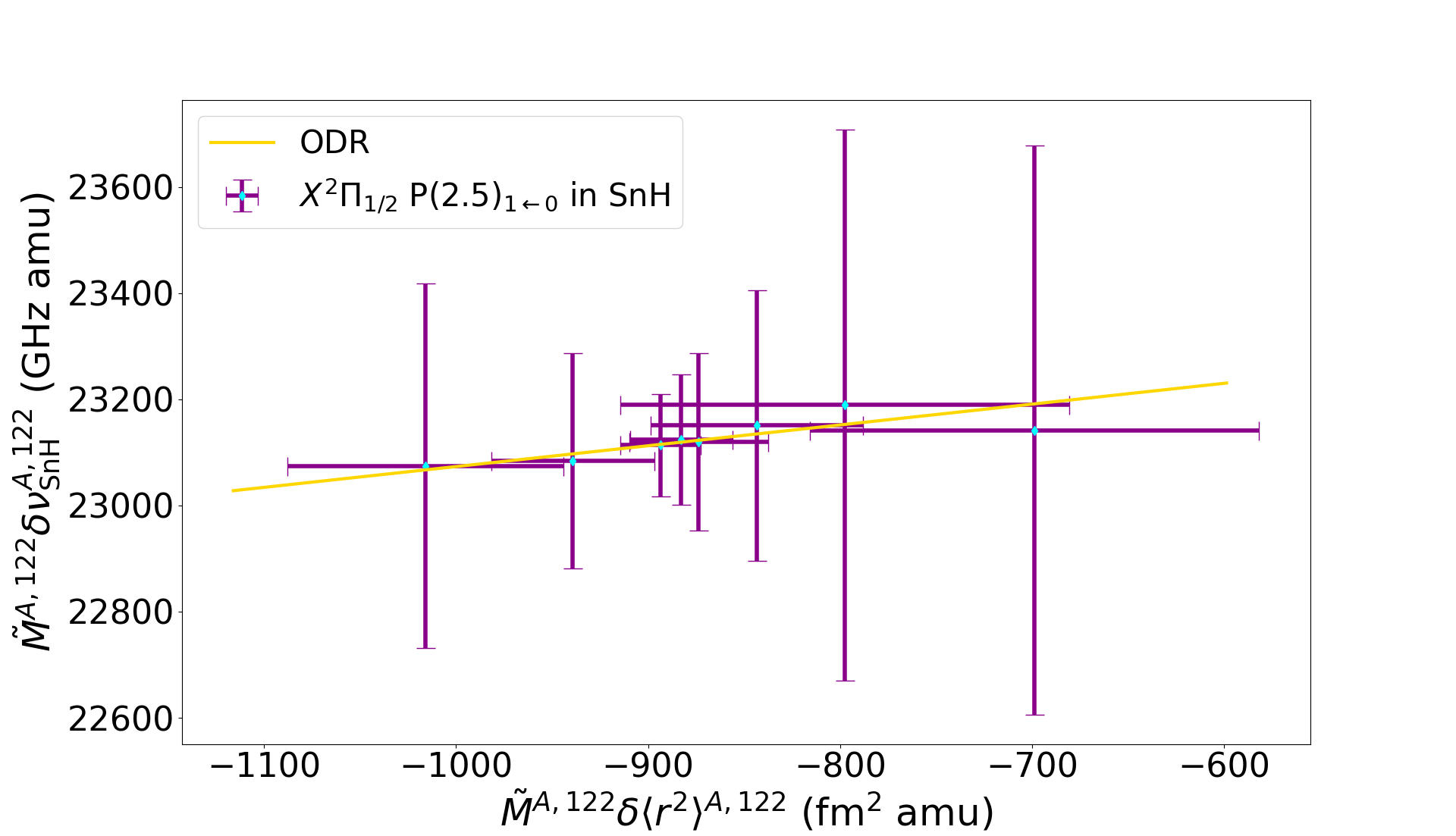}
    \caption{\small Mass-modified isotope shifts in the $P$(2.5)$_{1 \leftarrow 0}$ rovibrational transition in SnH against the mass-modified changes in mean-squared nuclear charge radii of Sn. All error bars arise from the 1$\sigma$ uncertainties of the literature isotope shifts (Table~\ref{tbl:snh}) and the uncertainties of the mass measurements taken from the 2020 Atomic Mass Evaluation~\cite{AME2020}.}
    \label{fig:snh_king_plot}
\end{figure}

Simon \textit{et al.} measured the low-lying rovibrational spectra (vibrational-rotational transitions within the $X^2\Pi_{1/2}$ electronic ground state) of the stable isotopomers of SnH using infrared laser spectroscopy~\cite{Simon1990}. Fig.~\ref{fig:snh_king_plot} shows a plot of the mass-modified isotope shifts in the $P$(2.5)$_{1 \leftarrow 0}$ rovibrational transition against the mass-modified changes in mean-squared nuclear charge radii of Sn from the analysis of the Barrett radii and the ratios of the radial moments obtained from non-optical experiments~\cite{FrickeHeiligSn}. This approach is fully equivalent to the King-plot analysis utilizing atomic isotope shifts, since the latter are linearly related to the changes in mean-squared nuclear charge radii. Plotting the mass-modified molecular isotope shifts against the mass-modified changes in radii allows to also test the validity of Eqn.~\ref{eqn:isotopeshift_final} using values of $\delta \langle r^2 \rangle^{A,A^{\prime}}$ taken from direct measurements of $\langle r^2 \rangle$ from non-optical methods. The tin chain provides a good test case for this alternative approach to the King-plot analysis, as it contains the highest number of stable isotopes across the periodic table, with a mass difference of more than 10 amu between the heaviest and lightest stable isotopes.

The transition frequencies along with the isotope shifts and the extracted changes in the mean-squared charge radii are listed in Table~\ref{tbl:snh}. A precision in the order of 38~MHz was reported in Ref.~\cite{Simon1990}, without giving an exact value. Therefore, a conservative uncertainty of 50~MHz was used for the molecular frequency measurements in the present analysis, leading to a slightly greater uncertainty in the isotope-shift values.

\begin{table*}[]
\renewcommand{\arraystretch}{1.65}
\caption{Transition frequencies (in cm$^{-1}$) for the $P$(2.5)$_{1 \leftarrow 0}$ rovibrational transition in the $X^2\Pi_{1/2}$ electronic ground state of SnH, and the corresponding isotope shifts (in GHz) with respect to $^{122}$SnH. The model-independent Barrett radii ($R^{\mu}_{k \alpha}$, in fm) and the ratios of the second radial moment ($V_2$) from Ref.~\cite{FrickeHeiligSn} used to calculate the literature values of the changes in mean-squared charge radii of Sn ($\delta \langle r^2 \rangle^{A,122}$ in fm$^{2}$) are given for completeness. The extracted changes in mean-squared charge radii of Sn from the isotope shifts in the $X^2\Pi_{1/2}$~$P$(2.5)$_{1 \leftarrow 0}$ transition in SnH ($\delta \langle r^2 \rangle ^{A,122}_{P\mathrm{(2.5)}}$ in fm$^2$) are also given.}
\label{tbl:snh}
\begin{tabular}{ccccccc}
\hline \hline
A   & $P$(2.5)$_{1 \leftarrow 0}$ & $\delta \nu _{P\mathrm{(2.5)}}^{A,122}$ & $R^{\mu}_{k \alpha}$ & $V_2$   & $\delta \langle r^2 \rangle ^{A,122}$ & $\delta \langle r^2 \rangle ^{A,122}_{P\mathrm{(2.5)}}$ \\ \hline 
112 & 1629.2956                     & +16.941(71)                                                                                                   & 5.8770(15)           & 1.27919 & -0.655(15)                            & -0.658(216)                                             \\
114 & 1629.1749                     & +13.323(71)                                                                                                   & 5.8979(15)           & 1.27931 & -0.509(15)                            & -0.502(207)                                             \\
116 & 1629.0581                     & +9.821(71)                                                                                                    & 5.9188(15)           & 1.27971 & -0.371(15)                            & -0.374(194)                                             \\
117 & 1629.0007                     & +8.100(71)                                                                                                    & 5.9250(14)           & 1.27981 & -0.330(15)                            & -0.341(223)                                             \\
118 & 1628.9455                     & +6.446(71)                                                                                                    & 5.9386(15)           & 1.27992 & -0.235(15)                            & -0.223(211)                                             \\
119 & 1628.8898                     & +4.776(71)                                                                                                    & 5.9423(14)           & 1.27999 & -0.210(15)                            & -0.206(205)                                             \\
120 & 1628.8364                     & +3.175(71)                                                                                                    & 5.9566(16)           & 1.28007 & -0.109(16)                            & -0.096(209)                                             \\
122 & 1628.7305                     & 0(0)                                                                                                         & 5.9723(15)           & 1.28022 & 0(0)                                  & 0(0)                                                    \\
124 & 1628.6282                     & -3.067(71)                                                                                                   & 5.9857(15)           & 1.28037 & +0.093(16)                           & +0.109(186)            \\ \hline \hline                                  
\end{tabular}
\end{table*}

If Eqn.~\ref{eqn:isotopeshift_final} holds for the rovibrational transition under consideration, a linear fit can be applied to the plot in Fig.~\ref{fig:snh_king_plot} whose slope corresponds to the field-shift factor~$\Psi$ and the y-intercept to the mass-shift factor~$\Xi$:
\begin{eqnarray}
    \nonumber
    \Psi_{P\mathrm{(2.5)}} = +0.4(1.6) \\
    \nonumber
    \Xi_{P\mathrm{(2.5)}} = +23470(1410)
\end{eqnarray}

The field-shift factor~$\Psi_{P\mathrm{(2.5)}}$ is consistent with 0, which implies that the experimental precision is not sufficient to unambiguously resolve the small field shift in the rovibrational transition. A small field shift is indeed expected for such a transition, as it does not involve a direct change of electronic state; the electronic overlap with the nucleus can change between the upper and lower rovibrational states only as a result of a coupling between the molecular rotation and the average electrostatic potential experienced by the electron.

As seen in Table~\ref{tbl:snh}, the nominal values of $\delta \langle r^2 \rangle^{A,A^{\prime}}_\mathrm{Sn}$ extracted from the molecular isotope shifts are in agreement with the values from non-optical experiments, falling within the error of the non-optical values for all isotopes. However, the values extracted from the molecular measurements are accompanied by large uncertainties, which originate from the small field shift of the transition and its large fractional uncertainty.

\section{\label{sec:discussion_outlook}Discussion and outlook}
As it can be seen from the results of Section~\ref{sec:analysis}, the changes in the mean-squared nuclear charge radii of Yb extracted using the 
% $^O P_{12}$(3) ($F^{\prime}=2$, $F^{\prime\prime}=2$)$_{A,0\leftarrow X,0}$ 
$T_0$ fitted parameter and the $^{O}P$-branch features in YbF are remarkably consistent with those extracted from the high-resolution spectroscopy of Yb$^{+}$. Similarly, the analysis of isotope shifts in ZrO and Zr$^{+}$ leads to consistent values for the changes in mean-squared charge radii between the zirconium ionic and the molecular systems.

In the case of SnH, while the nominal values of $\delta \langle r^2 \rangle^{A,A^{\prime}}_{\mathrm{Sn}}$ extracted from the infrared isotope shifts are in general agreement with the literature values from non-optical experiments, the uncertainties in the charge radii demonstrate that the field shift in the considered transition is consistent with 0. Possibly, had the experimental fractional uncertainty in the molecular measurements been smaller, the small but possibly non-zero field shift could have been resolved for the rovibrational transition. However, narrow-linewidth laser systems operating in the infrared regime are not widely available as of yet. Therefore, the present case might be generally representative of rovibrational transitions and their sensitivity to the field shift using state-of-the-art equipment. This result can guide experimentalists in selecting transitions to study nuclear-size effects in diatomic molecules. Additionally, as further discussed in Section~\ref{subsec:mol_struc_info}, the observation that the field shift in the vibrational transition is consistent with zero provides information on the change in the electronic wavefunction and its coupling to the nuclear motion.

Overall, the results of Section~\ref{sec:analysis} indicate that the laser spectroscopy of diatomic molecules can be used to accurately extract the changes in mean-squared nuclear charge radii consistent with the results from atomic (ionic) spectroscopy, as demonstrated with optical isotopes shift in YbF and ZrO. The analysis of the isotope shifts in the $T_0$ parameter in YbF, which includes values for isotopomers with odd-\textit{A} isotopes of Yb, also confirms that nuclear-structure effects, such as the appearance of odd-even staggering, can be captured by molecular measurements~(Fig.~\ref{fig:yb_ybf_radii}).

Furthermore, the results of the analysis of both $P$ and $R$ rotational lines with low and higher $J$ values alike in ZrO indicate that the molecular King-plot approach might also be a tool for molecular-structure studies. According to Eqn.~\ref{eqn:isotopeshift_final}, the molecular isotope-shift factors $\Psi$ and $\Xi$ are linked to the fundamental molecular parameters $f^{\Lambda}_{kl}$ and $\delta^{\Lambda}_{kl}$ that, analogous to the atomic case, provide information about the overlap of the valence electrons with the nuclear volume, and the breakdown of the Born-Oppenheimer approximation, respectively.

Additionally, as the values of $\delta \langle r^2 \rangle ^{A,172}_{\mathrm{Yb}}$ extracted from the $A ^2 \Pi_{1/2}$~$(\nu=0)$~$T_0$ parameter in YbF are consistent with those from the 369.4-nm in Yb$^{+}$, it is seen that the isotope shifts in the term energy~$T_0$ of a vibronic state can also be used in a King-plot analysis with accuracy. Experimental term-energy measurements in diatomic molecules are also possible with low-resolution laser spectroscopy (laser linewidth of a few GHz), as already demonstrated in the case of RaF~\cite{Udrescu2021}\footnote{In such cases, the uncertainty in the extracted $\delta \langle r^2 \rangle ^{A^{\prime}, A^{\prime\prime}}$ will probably be greater than the error bars for $T_0$ in Section~\ref{sec:analysis}, as the $T_0$ values used in the present analysis were extracted from a statistical fit involving high-resolution optical measurements.}. While the mass shift was not considered in the analysis of isotope shifts in Ref.~\cite{Udrescu2021} (due to the large mass of the species, which significantly dampens the presence of the mass shift, as seen in Eqn.~\ref{eqn:atomic_is}), the results of Section~\ref{sec:analysis} indicate that a complete consideration of the mass shift could be included using the King-plot method.
%
% In the case of pure rotational transitions, the field shift is small, as expected. Due to the inherently precise frequency measurements possible for such transitions, however, accurate values for the changes in the mean-squared charge radii can be extracted. The analysis of the pure rotational transitions in TiO leads to a different sign for the field shift and a slight change in the mass shift between the two transitions, which have different values of $J^{\prime\prime}$ (one with even value, one with odd) and $\Omega^{\prime\prime}$. A systematic analysis of the isotope-shift factors for pure rotational transitions with different quantum numbers can be undertaken to study the correlation between the rotational quantum numbers and the change in the electronic overlap with the nuclear volume.

The potential impact of the results of the present work can be discussed in relation to nuclear-structure studies with laser spectroscopy at RIB facilities and molecular-physics research.
\subsection{Importance for nuclear physics} \label{subsec:importance_nuclear_physics}
To trace how nuclear structure evolves across long isotopic and isotonic chains, RIB facilities have been developed, where nuclei with extreme proton-to-neutron ratios are produced for experimental study. Isotope separation on-line (ISOL) is a RIB production method that provides since many decades high-quality radioactive beams for precision nuclear structure studies and as probes for materials science, biochemical, and atomic research ~\cite{Blumenfeld2013,Borge2017}. With the thick-target ISOL technique, short-lived nuclides are produced by impinging highly energetic beams of protons onto a thick target material, typically composed of UC$_x$, LaC$_x$, UO$_2$, or ZrO$_2$, although other materials have also been successfully used. By heating the target material to a high temperature ($\sim$2000~$^\circ$C), the produced nuclei diffuse to the surface of the target as they form, and then effuse within a vapor into an ion-source region. Following ionization, the beam is mass-purified using magnetic dipolar mass separators, before being delivered to an experimental station. Overall, radioactive beams with half-lives down to a few milliseconds can be studied experimentally at RIB facilities.

Due to its versatility, laser spectroscopy -- collinear~\cite{Neugart2017} and in-source~\cite{Marsh2013,Heinke2016,Pohjalainen2016} -- has been used to study the evolution of ground-state properties across the chart of the nuclides, from helium ($Z=2$) isotopes up to the superheavies ($Z\geq 100$)~\cite{Campbell2016}. However, the area of the nuclide chart that has been studied with laser spectroscopy features several noticeable gaps, including the light and reactive isotopic chains from boron to fluorine and from silicon to chlorine ($Z<20$), the majority of refractory isotopic chains belonging to the transition metals, and most nuclides in the actinide region~\cite{Campbell2016}. The reason for these gaps is the reactive and/or refractory nature of the atoms: in the course of their diffusion and effusion from the ISOL target, reactive atoms form molecules with high efficiency, while refractive atoms stick to the target matrix. As a result, the delivered radioactive beams are of prohibitively limited intensity, far below the minimum requirements of state-of-the-art laser spectroscopy experiments ($\sim$0.1-100~atoms per second, depending on the technique).

A potential pathway towards the study of such nuclei with laser spectroscopy is to deliver them in the form of a molecular beam. Molecular extraction at ISOL facilities has already demonstrated improvements in the delivery of certain reactive~\cite{Ballof2019} and refractory~\cite{Koester2007} species. By exposing the ISOL target to a gas of choice during nuclide production, such as a fluorination or sulfurization agent, the species of interest can form molecules that are significantly more volatile than their atomic constituents.

The results of the present work demonstrate that the isotope shifts in radioactive molecules could be used to accurately extract values for $\delta \langle r^2 \rangle^{A,A'}$. 
% The molecules explored in this work were used as first test cases due to their simple electronic structure, having $\Sigma$ ground states with few valence electrons, but do not restrict the applicability of the approach to the presented cases only.
Importantly, as most refractory isotopic chains contain 3 or more stable isotopes, existing studies of isotope shifts (and isotope-shift factors) based on atomic laser spectroscopy and non-optical methods already exist. Therefore, the developed King-plot analysis for molecular measurements in the current work demonstrates that mean-squared charge radii of short-lived isotopes can be extracted from molecular isotope-shift measurements without the need for detailed molecular field- and mass-shift calculations. The analysis of ZrO/Zr$^{+}$ in Section~\ref{subsec:zro} is one such example; while in-target simulations show that a long range of radioactive Zr isotopes can be produced within a thick ISOL target, atomic Zr is refractory and therefore it cannot be extracted efficiently from a thick target. However, thin-ISOL approaches such as the one used in Ref.~\cite{Campbell2002} allow for the efficient extraction of atomic Zr, but the isotopic range producible with this technique is significantly more limited than for thick-ISOL techniques. Therefore, studying nuclear size effects in (volatile) molecular beams of Zr may allow for the study of short-lived Zr isotopes that are otherwise inaccessible with atomic spectroscopy. As Eqn.~\ref{eqn:isotopeshift_final} does not introduce limitations in terms of the non-substituted atom in the diatomic molecule, such a study can be done on fluorides, oxides, nitrides, etc., depending on the species that can be extracted most efficiently in each case. While this approach could be beneficial for nuclear-structure studies of the Zr chain, it is not the only case that would benefit from the approach; numerous reactive and refractory species have been successfully produced in molecular form, often with a considerable enhancement in beam purity or extraction efficiency~\cite{Au2022zenodo,Ballof2021Thesis,Koester2008}.

While, in the absence of a framework like the King-plot analysis, the need for detailed molecular-structure calculations prior to each experimental study would decelerate experimental progress, the ability to calculate these factors for diatomic molecules using quantum-chemistry methods is also important. As seen in Eqn.~\ref{eqn:isotopeshift_final}, the molecular isotope-shift factors $\Psi$ and $\Xi$ are linked to the molecular parameters $f_{kl}^{\Lambda,A}$ and $\delta_{kl}^{\Lambda,A}$ which could be calculated using \textit{ab initio} methods. In certain cases, such calculations can be more accurate in the diatomic molecule than in the constituent atomic systems, thus leading to more isotopic chains accessible by laser spectroscopy. The reason is that the laser-spectroscopic study of nuclei whose atomic structure contains more than two valence electrons, such as many lanthanide and actinide species, is often challenging. The valence electronic space of such atoms complicates the accurate calculation of the isotope-shift factors~\cite{Cheal2012}, which, in certain cases, can be very challenging even with state-of-the-art computational tools. Typically, such systems are studied as atomic ions, where the reduction in the valence electronic space allows for more accurate calculations. However, this strategy can lead to unfavorable transitions outside the optical range, or require high charge numbers that are beyond the technical capabilities of most RIB facilities. By placing the atom in a chemical bond, the number of valence quasiparticles included in the calculation can, strictly speaking, be reduced. As a result, the diatomic molecule can be within reach of computational packages such as implementations of the accurate and precise Fock-space relativistic coupled cluster (FS-RCC), which are currently limited to two valence quasiparticles~\cite{Oleynichenko2020}, and the extension of which to more valence quasiparticles is non-trivial. On the contrary, the constituent atom having three or more valence electrons would be incompatible with the same packages. For instance,  accurate relativistic coupled-cluster calculations of HfF$^+$~\cite{Skripnikov2017} and ThO~\cite{Skripnikov2016} have been successfully performed at the singles-doubles-triples and perturbatively quadruples level (CSSDT(Q)), while the Hf and Th atoms have four valence electrons and would therefore present a serious challenge for calculations at the same level of accuracy (multiple charge states of the Hf were recently calculated~\cite{Allehabi2021} only at the singles and doubles level, and the agreement with experiment was significantly worse than what is achievable with the approach in Refs.~\cite{Skripnikov2016, Skripnikov2017} as shown in the case of ThO). By utilizing a King-plot analysis that involves atomic and molecular isotope shifts, the atomic isotope-shift factors can be extracted from the molecular ones. The radioactive nuclei can then be studied within whichever system, atomic or molecular, leads to more favorable experimental parameters, such as efficiency and technical readiness.

Overall, Eqn.~\ref{eqn:isotopeshift_final} allows for relating the isotope-shift factors of atoms and molecules, and the link between the factors $\Psi$ and $\Xi$ and fundamental molecular parameters also allows for calculating the molecular isotope-shift factors with quantum chemistry. Consequently, the experimental progress in the laser spectroscopy of radioactive molecules can utilize decades of advances in atomic laser spectroscopy for cases where that is possible, as well as enable the study of systems that are currently inaccessible in atomic form due to technical or computational difficulties. Therefore, accelerated progress in the study of the charge radii of short-lived nuclei using laser spectroscopy could be achieved through the study of radioactive molecules. As \textit{ab initio} nuclear theory is progressing rapidly, expanding the experimental capabilities to more weakly bound isotopes and new isotopic chains is becoming progressively more important. In certain cases, joint experimental-theoretical studies of the nuclear charge radius of isotopes and chains that are currently experimentally inaccessible would provide a pathway to currently insufficiently studied physics, such as the role of many-body nuclear forces, continuum effects, nucleon clustering, and nucleon superfluidity, as aforementioned.

\subsection{Molecular-structure information} \label{subsec:mol_struc_info}
A close observation of the extracted isotope-shift factors for YbF, ZrO, and SnH in Section~\ref{sec:analysis} shows that a systematic study of $\Psi$ and $\Xi$ in a diatomic molecule could provide information about the electronic wavefunction and its coupling to the nuclear motion.

As seen from the isotope-shift factors of the rotational lines in ZrO (Table~\ref{tbl:zr_isotope_shift_factors}), while $\Psi$ and $\Xi$ are very similar for the lines $P$(1) and $R$(1), the mass-shift factor $\Xi$ scales very strongly with $J$ for the $P$ branch, while for the $R$ branch, both $\Psi$ and $\Xi$ scale strongly with $J$. Such patterns could be potentially used to benchmark theoretical calculations of the influence of the rotational quantum number~$J$ on the electronic overlap with the nucleus. Additionally, understanding the different behavior of the $P$ and $R$ lines would lead to molecular structure information that is not easily accessible through other experimental observables. For instance, $P$ and $R$ lines with a common rotational substate (such as $P$(1) and $R$(1)) could be used to extract the nuclear-electronic overlap and electron correlation energy in a single electronic state through the combination difference~\cite{Herzberg1945}, rather than a property that depends on the difference between two electronic states, such as the atomic case of \textit{F}.

In analogy with the atomic mass shift, the evolution of the molecular mass-shift factor $\Xi$ could provide information on the difference in electronic correlation energy between molecular states. In the molecular case, however, the normal mass shift, whose formulation is well-understood in atoms as the component of the mass shift that does not depend on electron correlations, might possess a different formulation with additional dependence on the vibrational and rotational degrees of freedom. For instance, in the case of YbF, the factors for the two \textit{P}-branch features exhibit an interesting effect: the mass-shift factor~$\Xi$ changes sign as \textit{J} increases, going from $+1720(240)$ for $J=3$ to $-4900(600)$ for $J=9$. While such a change might indicate that the coupling between molecular rotation and electronic energy leads to a drastic change in the electronic correlation energy for different rotational states, this interpretation is contradicted by the significantly weaker dependence of $\Psi$ on \textit{J}, and further unsupported by the very simple electronic structure of the free radical YbF. A possible explanation for the change of sign in $\Xi$ with increasing \textit{J} is that the molecular normal mass shift in fact also contains a strong dependence on the rotational quantum number \textit{J}. Systematic studies with the molecular King-plot method could thus potentially be used to study both the electronic correlation energy in molecules, even resolving its dependence on the vibrational and rotational degrees of freedom, and the contributions to the isotope shift that do not stem from electron correlations.

In the case of SnH, interestingly both the absolute and the relative magnitudes of the infrared isotope shifts (Table~\ref{tbl:snh}) are significantly greater than those typically encountered in atomic Sn (see for example Ref.~\cite{Gorges2019}). Since the field shift contributes only slightly, if at all (the contribution certainly is smaller than the resolution achieved in this analysis), isotope shifts of such magnitude for this transition might indicate that a molecular effect akin to the specific mass shift in atoms has a significantly larger magnitude in some diatomic molecules compared to their constituent atoms. As the specific mass shift originates from the change in the electron correlation energy as a result of the change in nuclear mass, molecular systems might be more sensitive probes for the study of such effects than their atomic constituents.

Additionally, although the field shift in the studied vibrational transition in SnH is consistent with zero, resolving a non-zero field shift in broadband vibrational transitions would also probe the electronic wavefunction, as the electrons spend less time on average within the nuclear volume(s) for a higher molecular vibrational energy. By tracing the evolution of the field-shift factor across transitions between states with high vibrational quantum numbers as well as overtone transitions ($\lvert \delta \nu \rvert > 1$), the nuclear-electronic temporal overlap can also be studied.

Understanding how the isotope-shift factors scale with the rotational (and vibrational, although not explicitly demonstrated in this work) quantum numbers is also of importance for mass-independent, multi-isotopomer studies for astrophysics. The detection of signatures of pure-rotational transitions in diatomic molecules has been recognized as a more powerful pathway to identify the isotopic composition of astrophysical environments compared to X-ray and $\gamma$-ray spectroscopy~\cite{McGuire2022}. However, for isotopomers containing short-lived radioactive atoms, the laboratory measurement of the rotational transition energies can be difficult, as it requires the development of equipment in specialized facilities that is not yet in place. Therefore, a mass-independent, multi-isotopomer analysis of the stable isotopomers of the molecules can be employed, to allow for the transition frequencies in the radioactive isotopomers to be extrapolated from those in the stable ones (for example, see Ref.~\cite{Breier2019}). In such an analysis, the field- and mass-related parameters ${f_{kl}}$ and ${\delta_{kl}}$ introduced in the Dunham expansion (Eqn.~\ref{eqn:dunhamfull}) are assumed to be independent of $\nu$ and $J$ for a given electronic state; that is, all vibrational and rotational transitions within a given electronic state are characterized by the same set of ${f_{kl}}$ and ${\delta_{kl}}$. The results of this work, however, demonstrate that this assumption is not fully realistic, since the field- and mass-shift factors $\Psi$ and $\Xi$, which are linked to ${f_{kl}}$ and ${\delta_{kl}}$ as per Eqn.~\ref{eqn:isotopeshift_final}, can be strongly dependent on $\nu$ and $J$. While extending the molecular King-plot analysis to pure-rotational transitions in diatomic molecules is beyond the scope of this work, further developments in this direction could therefore benefit the multi-isotopomer analysis for nuclear astrophysics.

To the knowledge of the authors, such molecular-structure information is not easily accessible by other techniques. The molecular King-plot analysis, as demonstrated in this work, opens a path for the study of new observables in molecular physics with laser spectroscopy. Importantly, laser-spectroscopic experiments on molecules have also been successful, as in the case of RaF~\cite{GarciaRuiz2020,Udrescu2021}, with ultra-high-sensitivity experiments, such as the collinear resonance ionization spectroscopy (CRIS) beamline, which is optimized to study species that remain bound for down to 5~ms~\cite{FarooqSmith2016} and rare species with production rates as low as a few tens per second~\cite{deGroote2017}. Therefore, field- and mass-shift studies can also be performed in weakly bound diatomic molecules, molecules containing short-lived radioactive nuclides, or molecular beams of low purity, making the approach versatile.

\section{\label{sec:conclusion}Conclusions}
In the present work, an isotope-shift expression for diatomic molecules that is linear with respect to the change in the nuclear charge radii~$\delta \langle r^2 \rangle^{A^{\prime},A^{\prime\prime}}$ was used in a King-plot analysis in combination with isotope-shift measurements in ionic atoms. Such approach can be used to extract $\delta \langle r^2 \rangle^{A^{\prime},A^{\prime\prime}}$ from molecular isotope shifts using known isotope shifts and isotope-shift factors in atomic transitions, or to study aspects of the molecular wavefunction.

To demonstrate the validity of the expression, a King-plot analysis of isotope shifts in the electronic spectra of $^{170-174,176}$YbF~\cite{Steimle2007} was performed in combination with isotope-shift measurements in the 369.4-nm transition in Yb$^{+}$~\cite{martenssonpendrill1994}.  The isotope shifts in both the fitted $A ^2 \Pi_{1/2}$~$(\nu=0)$~$T_0$ parameters of $^{170-174,176}$YbF as well as the $^O P_{12}$(3) ($F^\prime=2,$ $F^{\prime\prime} = 2$) and $^O P_{12}$(9) ($F^\prime=8,$ $F^{\prime\prime} = 9$) branch features of the $A ^2 \Pi_{1/2}$~$(\nu=0) \leftarrow X ^2 \Sigma^{+}$~$(\nu=0)$ transition of the isotopomers containing the even-$A$ Yb isotopes were analyzed.

The King-plot analysis was used to compare the values of $\delta \langle r^2 \rangle ^{A^{\prime},A^{\prime\prime}}$ extracted from the spectroscopy of YbF with those extracted using Yb$^{+}$. The agreement between the molecular- and atomic-extracted values is excellent, demonstrating that a King-plot analysis of optical molecular isotope shifts can be used with accuracy. The agreement in $\delta \langle r^2 \rangle ^{A^{\prime},A^{\prime\prime}}$ extracted both from the term energy as well as two \textit{P}-branch features indicates that both low- and high-resolution laser spectroscopy of molecules could be used for measurements of $\delta \langle r^2 \rangle ^{A^{\prime},A^{\prime\prime}}$.

Similarly, the analysis of optical isotope shifts in $P$- and $R$-branch rotational lines in the $C{}^{1}\Sigma^{+}-X{}^{1}\Sigma^{+}$ transition in ZrO with isotope shifts in the 327-nm transition in Zr$^{+}$ also yielded consistent values of $\delta \langle r^2 \rangle ^{A^{\prime},A^{\prime\prime}}$. Additionally, the difference in the isotope-shift factors between the $P$- and $R$-branch transitions for $J=1$ and $J=20$ reveals a quantum-number dependence on the change in the field and mass shift across rotational transitions. As a result, the King-plot method could be used to gain insights into molecular structure through a detailed study of the evolution of the isotope-shift factors as a function of the molecular quantum numbers.

Lastly, isotope-shift measurements of the $P$(2.5)$_{1 \leftarrow 0}$ rovibrational transition in the electronic ground state $X^2\Pi_{1/2}$ for the stable isotopomers of SnH~\cite{Simon1990} using infrared spectroscopy were analyzed in combination with values for the changes in mean-squared nuclear charge radii from non-optical techniques~\cite{FrickeHeiligSn}. The extracted field shift for this vibrational transition is consistent with zero, indicating that the transition is not sensitive to the nuclear volume within the achieved resolution. While this transition would therefore not be useful for the measurement of nuclear charge radii, this observation provides information on the change in the electronic wavefunction in a transition that does not involve a change in electronic state. To the knowledge of the authors, such information is not experimentally accessible via other methods.

The results of the current work can lead to experimental progress in nuclear-structure research by contributing to the study of nuclei with refractory and reactive atoms at thick-ISOL facilities, which are important for the study of unique nuclear phenomena and their role in the nuclear charge radius, while the systematic analysis of isotope-shift factors across many transitions in a molecule could be used to extract molecular-structure information that is not accessible by other methods.

\begin{acknowledgments}
M.A.-K. and G.N. acknowledge financial support from FWO, as well as from the Excellence of Science (EOS) programme (nr. 40007501), and the KU Leuven project C14/22/104. S. G. W. was supported by U.S. D.O.E. grant number DE-SC0021176. A. A. B. gratefully acknowledges financial support from the Deutsche Forschungsgemeinschaft (DFG, German Research Foundation) -- project number 328,961,117 -- SFB 1319 ELCH. The authors would like to thank R.~Berger (University of Marburg), T.~Steimle (Arizona State University), R.~F.~Garcia Ruiz (Massachusetts Institute of Technology), E.~Tiemann (Leibniz Universit\"{a}t Hannover), and Luk\'{a}\v{s} Pa\v{s}teka (Comenius University) for useful discussions and Sonja Kujanp\"a\"a (Jyv\"askyl\"a University) for Figure 1.
\end{acknowledgments}

\newpage
\bibliography{references}% Produces the bibliography via BibTeX.

\end{document}